\documentclass[prd,nofootinbib,floats,aps,twocolumn,tightenlines,superscriptaddress,floatfix]{revtex4-1}

\usepackage{graphicx}
\usepackage{dcolumn}
\usepackage{bm}
\usepackage[breaklinks=true,colorlinks=true,linkcolor=blue,urlcolor=blue,citecolor=blue]{hyperref}

\usepackage{tikz}
\usepackage{comment}
\usepackage{epsf,epsfig,amsmath,amssymb,dsfont}
\usepackage{amsthm,mathrsfs} 
\usepackage{graphicx} 
\usepackage{multirow}
\usepackage{float}
\usepackage{colortbl}
\usepackage{float}
\usepackage{bm}
\usepackage{longtable}
\usepackage{array}
\usepackage{mathtools}
\usepackage[normalem]{ulem}

\allowdisplaybreaks

\newcommand{\integral}[3]{\int_{#2}^{#3} \!\! \mathrm{d} #1 \,}

\newcommand{\ii}{\mathrm{i}}
\newcommand{\ee}[1]{\mathrm{e}^{#1}} 

\newcommand{\tr}{\operatorname{Tr}}
\newcommand{\id}{\mathbb{I}}

\newcommand{\intHd}[1]{H_{\text{int},\,#1}}
\newcommand{\intH}{H_{\text{int}}}

\newcommand{\comm}[2]{\left[{#1},{#2}\right]}

\newcommand{\ket}[1]{\left| {#1} \right\rangle}
\newcommand{\bra}[1]{\left\langle {#1} \right|}
\newcommand{\braket}[2]{\left\langle \vphantom{#2} {#1}\left|\vphantom{#1}{#2}\right.\right\rangle}
\newcommand{\ketbra}[2]{\left| {#1}\vphantom{#2} \right\rangle\!\left\langle {#2}\vphantom{#1} \right|}
\newcommand{\nn}{\nonumber\\}

\newcommand{\mOgamma}{\mathcal{O}\left(\|\gamma_1\|^2\right)}

\usepackage{color}

\newcommand{\blu}{\color{blue}}

\renewcommand{\Im}{\ensuremath{\mathrm{Im}}}

\begin{document}

\title{Transmitting qubits through relativistic fields}
\author{Robert H. Jonsson}
\affiliation{Microtechnology and Nanoscience, MC2, Chalmers University of Technology, SE-412 96 G\"oteborg, Sweden}
\affiliation{QMATH, Department of Mathematical  Sciences,  University  of  Copenhagen,
Universitetsparken  5,  2100  Copenhagen,  Denmark}
\author{Katja Ried}%
\affiliation{Institute for Quantum Computing, University of Waterloo, Waterloo, Ontario, N2L 3G1, Canada}
\affiliation{Department of Physics \& Astronomy, University of Waterloo, Waterloo, Ontario, Canada, N2L 3G1}
\affiliation{Perimeter Institute for Theoretical Physics, 31 Caroline St N, Waterloo, Ontario, N2L 2Y5, Canada}
\affiliation{Institut f\"ur Theoretische Physik, Universit\"at Innsbruck, Technikerstra{\ss}e 21a, 6020 Innsbruck, Austria}
\author{Eduardo Mart\'{i}n-Mart\'{i}nez}
\affiliation{Institute for Quantum Computing, University of Waterloo, Waterloo, Ontario, N2L 3G1, Canada}
\affiliation{Department of Physics \& Astronomy, University of Waterloo, Waterloo, Ontario, Canada, N2L 3G1}
\affiliation{Perimeter Institute for Theoretical Physics, 31 Caroline St N, Waterloo, Ontario, N2L 2Y5, Canada}
\affiliation{Department of Applied Mathematics, University of Waterloo, Waterloo, Ontario, N2L 3G1, Canada}
\author{Achim Kempf}
\affiliation{Institute for Quantum Computing, University of Waterloo, Waterloo, Ontario, N2L 3G1, Canada}
\affiliation{Department of Physics \& Astronomy, University of Waterloo, Waterloo, Ontario, Canada, N2L 3G1}
\affiliation{Perimeter Institute for Theoretical Physics, 31 Caroline St N, Waterloo, Ontario, N2L 2Y5, Canada}
\affiliation{Department of Applied Mathematics, University of Waterloo, Waterloo, Ontario, N2L 3G1, Canada}

\begin{abstract}

Wireless communication derives its power from the simultaneous emission of signals in multiple directions. However, in the context of quantum communication, this phenomenon must be reconciled carefully with the no-cloning principle.
In this context, we here  study how wireless communication of quantum information can be realized via relativistic fields.  To this end, we extend existing frameworks to allow for a non-perturbative description of, e.g., quantum state transfer. We consider, in particular, the case of 1+1 spacetime dimensions, which already allows a number of interesting scenarios, pointing to, for example, new  methods for tasks similar to quantum secret sharing.
\end{abstract}

\maketitle

\section{Introduction}
Wireless communication, be it through fundamental fields such as the electromagnetic field in the vacuum or through the fields of collective degrees of freedom in condensed matter systems, can be a  powerful tool, allowing a sender to transmit information to multiple receivers without the need for much infrastructure.   
This raises the question of whether wireless communication can be used to transmit not only classical, but also quantum information. This is particularly interesting in light of the latest development in satellite-based quantum communication \cite{rideout_fundamental_2012,ren_ground-to-satellite_2017}, which may open up new  prospects for studying the impact of relativistic effects on quantum communication \cite{bradler_infinite_2009,bradler_private_2009,cliche_relativistic_2010,bradler_quantum_2012, downes_quantum_2013,bruschi_spacetime_2014,landulfo_nonperturbative_2016,jonsson_quantum_2017,gianfelici_quantum_2017}.

In the context of classical information, a key advantage of wireless communication is that it enables a sender to transmit to many receivers, in different directions. However, any attempt to transmit quantum information through a quantum field to many receivers faces a major obstacle in the form of the no-cloning principle.
Concretely, one can show that the channel from the sender to any single receiver in such a permutationally invariant scenario is anti-degradable, and consequently has zero quantum capacity
\footnote {Since this article focuses on fundamental questions concerning quantum fields as a medium for wireless quantum communication in relativistic scenarios, no other additional side-channels are assumed. Therefore, the article considers the quantum capacity $Q$, as first introduced by Lloyd \cite{lloyd1997capacity}. 
(See also \cite{kretschmann2004capacity} for a thorough discussion of many equivalent definitions found in the literature.) 
In particular, we preclude the additional possibilities offered by two-way classical communication (which are explored, e.g., in \cite{bennett1996entangQEC}). }
(see Appendix \ref{sec:symmzeroqc}.) {{\color{red}} There are two workarounds. A non-zero quantum capacity can be achieved when transmitting quantum information through a quantum field if the sender's signal is highly focused towards a single receiver.  
}
Alternatively, when two or more receivers receive  signals from the sender, these receivers can cooperate and thereby establish a channel with non-zero quantum capacity with the sender. 
The fact that the receivers need to cooperate is of interest because it may enable the implementation of tasks such as quantum bit commitment or quantum secret sharing 
\cite{kent_unconditionally_1999,hillery_quantum_1999,gottesman_theory_2000,adlam_device-independent_2015}. (See Appendix \ref{sec:qipwsymmetry}.)

In this context,  
the present Article studies the  transmission of quantum information in wireless communication between localized signaling devices, focusing, in particular, on the task of  \textit{quantum state transfer} \cite{cirac_quantum_1997}.

The conventional account of communication through, e.g., the electromagnetic field is straightforward: for example, an excited atom decays, emitting a photon, and another atom in its ground state absorbs the photon. However, a rigorous analysis in  quantum field theory shows that there are pitfalls that need to be carefully navigated regarding, in particular, often-used approximations that can introduce a subtle break of causality \cite{martin-martinez_causality_2015}. Important for our purposes here is the quantum channel between localized quantum systems, such as atoms, that communicate through a relativistic quantum field, a concept that was first introduced  in \cite{cliche_relativistic_2010}. It has been shown that such quantum channels exhibit surprising phenomena, such as the ability to transmit classical information without transmitting energy in certain circumstances \cite{jonsson_information_2015,jonsson_information_2016,jonsson_quantum_2017}. 

In the present work we propose a non-perturbative protocol for wireless quantum communication in this scenario. Going beyond previous work by Landulfo \cite{landulfo_nonperturbative_2016}, our protocol achieves non-zero quantum capacity. To this end, we use an extension of non-relativistic quantum state transfer protocols that were originally designed for electromagnetic fields in an optical cavity \cite{leghtas_deterministic_2013}.

Within the scope of this article we model the senders and receivers, i.e., the localized quantum systems, such as atoms, which couple to a quantum field, such as the electromegnetic field as Unruh-DeWitt particle detectors that are coupled to a massless scalar field in 1+1 dimensional spacetime. On one hand, working in 1+1 dimensions has the advantage that signals there can only propagate to the right or to the left, but do not dilute as  the distance from the emitter increases.

On the other hand, and more importantly, the scenario considered here closely resembles the effectively one-dimensional cavities and waveguides that are used for current implementations of quantum information processing using superconducting circuits \cite{blais_cavity_2004,mckay_finite_2017,forn-diaz_ultrastrong_2017}.  These technologies are entering regimes where the finite speed of propagation of light becomes relevant, and where, consequently, relativistic effects may impede or enable quantum information processing in novel ways. 

Within the setting described above, we develop a state transfer protocol between two atoms, as detailed  in Sections  \ref{sec:AliceCoupling} and \ref{sec:BobCoupling},  which achieves approximate quantum state transfer with arbitrary small error.
We further show, in Section \ref{sec:delocalizingcavity}, how a sender who couples  to both left- and right-moving momentum of the field can delocalize quantum information in the cavity field  such that a receiver can only access the message at specific focal points. 
We conclude with an outlook, addressing, among other points, how our protocols can be generalized to higher spacetime dimensions by coupling sender and receiver to field observables with narrow directional propagation profiles.

In higher spacetime dimensions, where scenarios with many receivers arise generically, the consequences of the no-cloning principle discussed above will naturally be of high importance.
Appendix \ref{sec:symmetriccouplingissues} therefore gives a review of quantum information theoretical notions useful for analysing such scenarios. 
In particular, Section \ref{sec:symmzeroqc} derives the anti-degradability of the quantum channel between sender and single receivers in permutationally invariant scenarios. Based on this, Section \ref{sec:qipwsymmetry}  discusses some examples of interesting quantum information processing tasks that may be possible despite -- or precisely because of -- the presence of symmetric signals in wireless communication.
 
Throughout the paper we use natural units, $c=\hbar=1$.

\section{Transmitting a qubit state into the field}\label{sec:AliceCoupling}

In the following, we develop a concrete model of quantum information transmission between local observers via relativistic quantum fields.
The impact of   relativistic and gravitational effects on the propagation of quantum information inside a relativistic quantum field was addressed in  \cite{downes_quantum_2013,bruschi_spacetime_2014}. The quantum optical communication protocols considered there are based on direct access of the sender and receiver to localized modes of the field.
In contrast, here we include the signaling devices into our framework. 
To this end, we model sender and receiver as local quantum systems that couple to the field via a unitary interaction.
This section begins with a brief review of the Unruh-DeWitt particle detector model that we use, followed by a discussion of  earlier studies of quantum information transmission between detectors.

The framework studied in this article is not only interesting from a fundamental point of view, but also provides a prototype model which can be used to explore novel methods of quantum information processing with  relativistic fields.
In particular the 1+1 dimensional scenarios on which this article focuses may be implementable in cavities or superconducting circuits, for example. These  have already been used to demonstrate relativistic effects like the Casimir effect, and recent experiments implemented ultra-strong and fast switchable couplings  \cite{blais_cavity_2004,johansson_dynamical_2010,wilson_observation_2011,mckay_finite_2017,forn-diaz_ultrastrong_2017}.

\subsection{Modelling the light-matter interaction with Unruh-DeWitt detectors}\label{sec:UDW}

We will model the interaction of the detectors operated by Alice and Bob by variations of the well-known Unruh-DeWitt model \cite{dewitt_quantum_1979} . Although simple, this detector model captures most of the fundamental features of the light-matter interaction when there is no exchange of angular momentum \cite{martin-martinez_wavepacket_2013,alhambra_casimir_2014,pozas-kerstjens_entanglement_2016}.

The Unruh-DeWitt detectors (from now on referred to as the `atoms' or `detectors') are two-level systems, with energy eigenstates $\ket{g}$ and $\ket{e}$, which interact with a background scalar field $\phi$. The interaction  Hamiltonian takes the general form 
\begin{align}
\intH(t)=\!\!\!\!\!\sum_{\nu\in\{\text{A,\,B}\}}\!\!\!\!\lambda_{\nu}\chi_{\nu}(t)m_{\nu}(t)\!\int\text{d}^n\bm{x}\,f_{\nu}(\bm{x}-\bm{x}_{\nu})\phi(\bm{x},t)
\label{eq:hamiltonian}
\end{align}
in the interaction picture.
Here $\nu\in\{\text{A},\text{B}\}$ labels Alice's and Bob's detectors, $\lambda_{\nu}$ is the overall coupling strength, and $\chi_\nu(t)$ is the switching function, which controls the interaction time  of each detector with the field. $m_{\nu}(t)$ is the monopole moment of each detector,
whose time-dependence in the interaction picture is given by
\begin{align}
m_{\nu}(t)=\ketbra{e}g_\nu e^{\ii\Omega_{\nu}t} +\ketbra{g}e_\nu e^{-\ii\Omega_{\nu}t}\label{eq:moment},
\end{align}
where $\Omega_\nu$ is the energy gap between the  stationary states of the detector.
Finally, $f_{\nu}(\bm{x})$ is the spatial profile of each detector, with $\bm x_\nu$  denoting the center-of-mass position. 
The interaction $\intH(t) $ can also couple the detector to other field observables than the amplitude $\phi(\bm{x},t)$. We will make use of this in the following, and consider detectors that couple, e.g., to the right-moving momentum of the field.

The time evolution under this coupling is commonly studied using perturbation theory.
However, perturbative methods are insufficient for the study of quantum information transmission, since (e.g., in quantum state transfer) 
the receiver may end up in a state orthogonal to their initial state, which clearly cannot be viewed as merely a perturbative change of state.

The  quantum capacity of the channel between two particle detectors has been addressed recently by Landulfo \cite{landulfo_nonperturbative_2016}.
There, for detectors with a vanishing energy gap, it was shown that the quantum channel is \emph{entanglement-breaking}, which implies zero quantum capacity.
(A channel is entanglement-breaking if it can be simulated by performing a measurement on the input state and transmitting only  the classical information about the outcome to the receiver.)

In fact, a closer analysis of the scenario in \cite{landulfo_nonperturbative_2016} shows that not only the channel from the sender to the receiver, but already the channel from the sender to the field is entanglement-breaking, meaning that only classical information is transmitted from the sender to the field in the first place.
This is due to the vanishing energy gap $\Omega_\nu=0$ of the detectors. For zero-gap detectors, the free detector Hamiltonian is effectively zero (formally, 
proportional to the identity):
as a result, the state of the field after interacting with Alice's detector depends only on the measurement outcome of  a single fixed observable -- the interaction Hamiltonian -- with respect to Alice's initial state.

One way to achieve non-zero quantum capacity would be to use detectors with a non-zero energy gap, since the non-trivial free Hamiltonian can be thought of as dynamically changing the observables while the detector is coupled to the field.  However, general solutions for this case have not yet been developed.

We will avoid this problem by instead allowing two instantaneous interactions, at different times $t=t_i$ and with different observables.
This idealization (see e.g.,~ \cite{hotta_quantum_2008,simidzija_non-perturbative_2017}) still admits  a straightforward non-perturbative treatment. 
With this  kind of coupling,  
if we choose two interaction Hamiltonians that do not commute, we can achieve quantum state transfer from the detector to the field.

\subsection{Instantaneous interaction yields a controlled displacement operator}

Using Unruh-DeWitt detectors, we will now construct a coupling that effectively  applies a displacement operator to the field, conditioned on the state of the detector. We assume the detector to be a two-level system and take the coupling to be localized at a single time $t=t_i$. 
For the remainder of this article, we assume the field to be  a massless scalar Klein-Gordon field in (1+1)-dimensional Min\-kow\-ski spacetime. (Section \ref{sec:delocalizingcavity}  considers a one-dimensional cavity.)

All couplings in our signalling protocol are of the same general form, which is a slight variation of the general Unruh-DeWitt interaction Hamiltonian introduced above.
We introduce and discuss it here, using  the example of the   Hamiltonian of Alice's first coupling to the field, which reads
\begin{align}\label{eq:Hint1}
\intHd{A}^{(1)}(t)= \mu_A\,\delta(t-t_0)  \frac12 \left(\id+\sigma_X\right)\otimes \integral{x}{}{} f(x) \pi_-(x,t).
\end{align}

We highlight the following differences from the general case:
The coupling constant that sets the overall strength of the interaction is now denoted
by $\mu_A$ (instead of $\lambda_A$). This reflects the fact that it has the dimension of mass, which ensures the correct dimension of the overall Hamiltonian.
The Dirac distribution $\delta(t-t_0)$ serves as switching function, modelling an instantaneous interaction  
at time $t=t_0$. 

For the purpose of constructing a basic protocol of quantum state transfer, a convenient choice of the  
detector observable is
\begin{align}
\frac12\left(\id+\sigma_X\right)=\ketbra{+X}{+X},
\end{align}
i.e., the projector onto the $+1$ eigenstate of $\sigma_X$, instead of the standard monopole operator in equation \eqref{eq:moment}.

Finally, $\integral{x}{}{} f(x) \pi_-(x,t_0)$ is the field observable to which the detector is coupled. 
The function $f(x)$ describes the spatial profile of the detector.  We choose it real-valued and compactly supported.
Weighted by this profile function, the detector couples to the right-moving part $\pi_-(x,t)$ of the conjugate momentum of the field. (In 1+1 spacetime dimensions, the conjugate momentum of the field $\pi(x,t)=\partial_t\phi(x,t)$ splits up into a right-moving and a left-moving part, $\pi=\pi_-+\pi_+$. More details on this  can be found, e.g., in \cite{jonsson_information_2016}.)
The right-moving momentum is given by
\begin{align}
\pi_-(x,t) &= \frac12\left(\partial_t\phi(x,t)-\partial_x\phi(x,t)\right)\nn
&=\integral{k}0\infty (-\ii)\sqrt{\frac{k}{4\pi}} \left( \ee{-\ii k (t- x)}a_{k}- \ee{\ii k (t- x)}a_{k}^\dagger \right).
\end{align}

By coupling the detector to a right-moving observable of the field, we ensure that all information about Alice's initial state only propagates in one direction. 
This overcomes the obstacles that would arise from a symmetric coupling to the field, which are  discussed in Appendix \ref{sec:symmetriccouplingissues}.

We will now show that the 
time evolution under $\intH^{(1)}(t)$ implements a conditional multi-mode coherent state displacement.
First, note how the instantaneous coupling allows straightforwardly for a non-perturbative treatment of the interaction by  eliminating the time ordering $\mathcal{T}$ from the Dyson series expansion of the time evolution operator. Therefore, the unitary relating the joint field-detector states before and after the interaction can be written as
\begin{align}
U^{(1)}&= \mathcal{T}\exp\left( -\ii \integral{t}{}{}\intH^{(1)}(t) \right) \nn
&= \exp\left( -\ii \mu_A\frac12 \left(\id+\sigma_X\right) \otimes \integral{x}{}{} f(x) \pi_-(x,t_0) \right)\nn
&= \ketbra{+X}{+X}\otimes \exp\left( -\ii \mu_A \integral{x}{}{} f(x) \pi_-(x,t_0) \right)\nn
&\qquad\qquad+\ketbra{-X}{-X}\otimes \id.
\end{align}

Notice that the field operator can be written as
\begin{align}
& -\ii \mu_A \integral{x}{}{} f(x) \pi_-(x,t_0) \nn 
&=  \mu_A \integral{x}{}{} f(x) \integral{k}0\infty \sqrt{\frac{k}{4\pi}} \left( \ee{\ii k (t_0- x)}a_{k}^\dagger - h.c. \right) \nn 
&= \integral{k}0\infty \left( \frac{\mu_A  \ee{\ii k t_0}\sqrt{k} }{2\sqrt\pi}\tilde{f}(k)a_{k}^\dagger - h.c. 
\right)
\end{align}
where we denote
\begin{align}\label{eq:profilefourier}
\tilde{f}(k)=\integral{x}{}{} f(x)\ee{\ii k x}.
\end{align}
Defining
\begin{align}\label{eq:DefDisplAlpha}
\alpha_1(k)=  \mu_A\ee{\ii k t_0} \sqrt{\frac{k}{4\pi}}  \tilde{f}(k) \theta(k),
\end{align}
where $\theta(k)$ denotes the Heaviside function, one can then write the time evolution operator as
\begin{align}\label{eq:ctrldisplop}
U^{(1)} &= U_{\alpha_1}^{(+X)}:= \ketbra{+X}{+X}\otimes D_{\alpha_1}+\ketbra{-X}{-X}\otimes \id
\end{align}
where $D_{\alpha_1}=\exp\left( \integral{k}{}{}\left[ \alpha_1(k) a_k^\dagger - \alpha_1(k)^* a_k\right]\right)$ is a continuous multi-mode displacement operator (see Appendix \ref{app:coherent}).
By the notation $ U_{\alpha_1}^{(+X)}$ we indicate that the operator displaces the field by $\alpha_1$ conditioned on Alice's detector being in the state $\ket{+X}$.

\subsection{Transferring Alice's state to the field}\label{sec:transferAlicetoField}
We will now show how one can transfer Alice's state to the field by
using two non-commuting  controlled displacement couplings.

We assume that Alice's detector and the field start out in the pure product state
\begin{align}\label{eq:initialAliceField}
\ket\zeta&=\ket\psi_A\ket0= \left(x_+\ket{+X}_A +x_-\ket{-X}_A\right)\ket0.
\end{align}
The first interaction $U^{(1)}$, discussed in the previous section, evolves this initial state  into 
\begin{align}\label{eq:secIIIafterU1}
U^{(1)}\ket\zeta&= U^{(+X)}_{\alpha_1}\ket\zeta= x_+ \ket{+X}_A\ket{\alpha_1}+ x_-\ket{-X}_A\ket0.
\end{align}
The partial state of the field after this first coupling is 
\begin{align}
\tr_A U^{(1)}\ketbra\zeta\zeta U^{(1)\dagger}=  \left|x_+\right|^2 \ketbra{\alpha_1}{\alpha_1}+\left|x_-\right|^2 \ketbra00.
\end{align}
Note that this is a function only of the expectation value $\bra\psi \sigma_X \ket{\psi}_A=\left|x_+\right|^2 - \left|x_-\right|^2=2\left|x_+\right|^2-1$.  
This implies that the channel from Alice's initial state to the field state is entanglement-breaking and thus has zero quantum capacity.

In order to enable the  transmission of quantum information to the field, we add a second interaction, at a later time $t=t_1$,  
which couples the detector through the $\sigma_Z$ Pauli matrix instead of $\sigma_X$ in \eqref{eq:Hint1}. 
This implies a unitary
\begin{align}\label{eq:U2}
U^{(2)}=U_{\alpha_2}^{(+Z)}=\ketbra{+Z}{+Z}\otimes D_{\alpha_2}+\ketbra{-Z}{-Z}\otimes\id
\end{align}
with the displacement
\begin{align}\label{eq:DefDisplBeta}
\alpha_2(k)=\mu_A\ee{\ii k t_1} \sqrt{\frac{k}{4\pi}} \tilde{f}(k)\theta(k).
\end{align}
This displacement differs from the first coupling only by a $k$-dependent phase, $\alpha_1(k)=\alpha_0(k) \ee{\ii k (t_1-t_0)}$, because we assume the detector profile to remain constant.

To simplify subsequent calculations, we assume that the time delay $t_1-t_0$ between the two interactions is larger than the maximal diameter of the support of the detector profile function $f(x)$. This means that the spacetime points at which the detector couples the second time lie inside the future lightcone of the spacetime points at which the detector interacts with the field the first time, at $t=t_0$.
Under this assumption, the two displacement operators $D_{\alpha_1}$ and $D_{\alpha_2}$ 
commute, as discussed in \eqref{eq:commD}:
\begin{align}
\comm{D_{\alpha_1}}{D_{\alpha_2}}=0.
\end{align}
This allows us to write the state after the second coupling as
\begin{align}\label{eq:stateafterAlice}
    U\ket\zeta &=U^{(2)} U^{(1)}\ket\zeta \nn
&= \frac1{\sqrt2} \left( \ket{+Z}_A \left( x_+ \ket{\alpha_2+\alpha_1}+x_-\ket{\alpha_2}\right)  \right.\nn* 
    &\qquad\qquad \left. +\ket{-Z}_A \left( x_+\ket{\alpha_1}-x_-\ket0\right) \right).
\end{align}

Note that Alice, independently of what her initial state was,  is (almost) maximally entangled with the field after the two couplings if the four field states $\{\ket0,\ket{\alpha_1},\ket{\alpha_2},\ket{\alpha_1+\alpha_2}\}$ are pairwise (almost) orthogonal, i.e., if their mutual overlap is neglible ($\braket{\alpha_1}{\alpha_2}\approx0$ etc.). 
This is the case when  $\integral{k}{}{}\left|\alpha_1(k)\right|^2$ and $\integral{k}{}{}\left|\alpha_2(k)\right|^2$ are large (see \eqref{eq:coherentstateproduct}), which can be achieved by choosing the coupling constant $\mu_A$ large enough.
However, despite Alice being maximally entangled with the field, a measurement of Alice's detector can only reveal \emph{in which way} her initial state (parametrized by the coefficients $x_\pm$) is now encoded in the field, but it cannot reveal information about that state itself. All information about Alice's initial state has been transferred to the field.

\subsection{Quantum capacity of the channel from Alice to the field}\label{sec:CapacityAliceToField}

In order to assess the quantum capacity of the channel from Alice's initial state to the state of the field after the second coupling, we now calculate a lower bound on it.
To this end, we introduce a fictitious ancillary qubit $A'$, which is initially maximally entangled with Alice's detector,
\begin{align}\label{eq:bellstateinitial}
\ket{\psi}_{AA'}=\frac1{\sqrt2} \left(\ket{+X}_A\ket{+X}_{A'}-\ket{-X}_A\ket{-X}_{A'}\right).
\end{align}
The coherent information $I(A'>F)$ between the ancilla and the field after the field-detector interaction has taken place then provides a lower bound to the capacity of the channel from input qubit $A$ to the final state of the field \cite{lloyd1997capacity, shor2002capacity, devetak2005private}.
This coherent information can be conveniently rewritten (see Appendix \ref{app:coherentinfo}) in terms of the marginal final state $\rho_{AA'}\equiv\tr_{F} U \ket\psi_{AA'}\ket0_F \bra\psi_{AA'}\bra0_F U^\dagger $ as
\begin{align}\label{eq:coherentinfoalicetofield_simple}
I(A'>F)&=
S\left(\rho_{AA'} \right)
- S\left( \tr_{A'} \rho_{AA'} \right),
\end{align}
eliminating the need to compute states of the infinite-dimensional field.

In order to obtain $\rho_{AA'}$, we note that the overall state after the two couplings is
\begin{align}
& U \ket{\psi}_{AA'}\ket0_F\nn* 
&\qquad =\frac1{2} \ket{+Z}_A \left( \ket{+X}_{A'} \ket{\alpha_2+\alpha_1}+ \ket{-X}_{A'}\ket{\alpha_2}\right)  \nn* 
    &\qquad\qquad  +\frac12
    \ket{-Z}_A \left( \ket{+X}_{A'}\ket{\alpha_1}-\ket{-X}_{A'}\ket0\right).
\end{align}
One can see immediately that, independently of the result of a hypothetical measurement on Alice's detector, the field and the ancilla end up in an entangled state, suggesting that coherence is preserved. 
In the limit where the overlap between the coherent field states $\{\ket{\alpha_1+\alpha_2},\ket{\alpha_1},\ket{\alpha_2},\ket0\}$ 
can be neglected, the marginal $\rho_{AA'}$ becomes maximally mixed, hence the coherent information is $I(A'>F)=1$.

\begin{figure}
\centering
\includegraphics[width=0.45\textwidth]{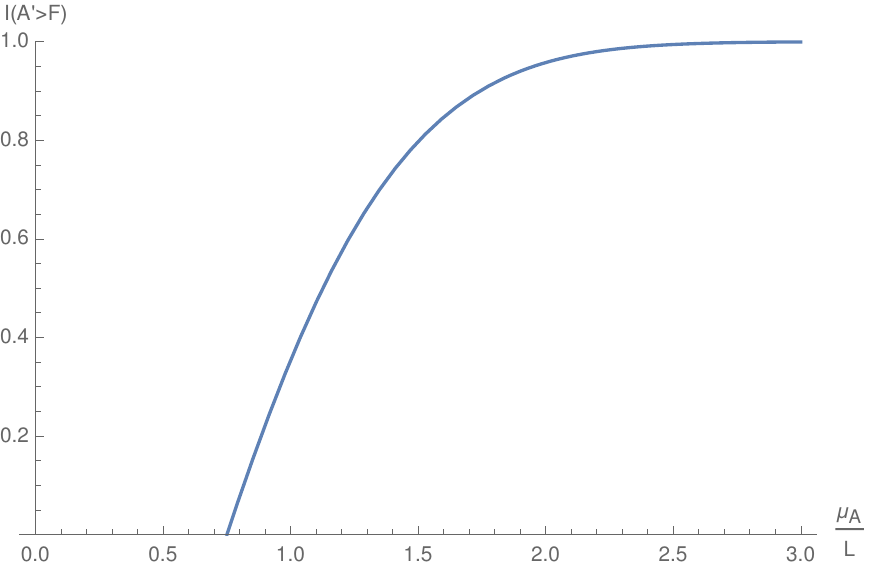}
  \caption{Lower bound on the coherent information $I(A'>F)$ in \eqref{eq:coherentinfoalicetofield} of the channel from Alice's detector to the field for a triangle-shaped detector profile \eqref{eq:triangledetectorprofile}. The lower bound is obtained by evaluating \eqref{eq:coherentinfoalicetofield_simple} for the initial state \eqref{eq:bellstateinitial}. As long as the separation between the two couplings is strictly timelike $t_1-t_0>L$, the influence of the time delay $t_1-t_0$ is negligible. (The plot  shows $t_1-t_0=1.5L$.) For strong enough couplings the coherent information soon approaches its maximum possible value of 1 bit. This means that all information about Alice's initial state is  transferred to the field.}
  \label{fig:coherentinfo} 
\end{figure}

In order to verify that the protocol yields non-zero quantum capacity from Alice to the field even when there is some overlap between the field states, we performed numerical calculations for a particular case. We chose the spatial profile of Alice's detector to be a triangle function, 
\begin{align}\label{eq:triangledetectorprofile}
f(x)=\frac2L \left(1-\frac{2|x|}L \right) \theta\left(\frac2L-|x|\right),
\end{align}
which has support on the interval $-L/2<x<L/2$ and $\integral{x}{}{}f(x)=1$.
The size of the displacement resulting from this profile function is given by
\begin{align}
    \|\alpha_1\|^2&=\integral{k}{}{} \left|\alpha_1(k)\right|^2 =\integral{k}{0}{\infty}  \frac{\mu_A^2 }{4\pi} k \left|\tilde{f}(k)\right|^2 \nn*
& = \frac{4\ln 2}{\pi}  \left(\frac{\mu_A}{L}\right)^2,
\end{align}
i.e., the overlap of the displaced field state $\ket{\alpha_1}$ and the vacuum state is $ \braket{\alpha_1}0=\ee{-\|\alpha_1\|^2/2}=4^{-(\mu_A/L)^2/\pi }$. Note that it is a function only of the ratio between the coupling constant $\mu_A$ and the detector diameter $L$.
Since we assume that the two interactions of Alice with the field are strictly timelike separated, we have $t_1-t_0>L$. Under this condition, the influence of the exact value $t_1-t_0$ on (the lower bound on) the coherent information is negligible. 

The results for the lower bound on the coherent information of the channel from the detector to the field, as a function of the coupling strength, are shown in Figure \ref{fig:coherentinfo}. 
One can see that, with increasing coupling strength, the coherent information - and thus the quantum capacity of the channel from Alice's detector to the field - approaches  1 bit, which is the maximum value possible. In particular, for coupling strenghts $\mu_A \gtrapprox 0.75 L$, Alice can transmit quantum information into the field with non-zero quantum capacity.

\section{Retrieving the qubit state from the Field}\label{sec:BobCoupling}

In the previous section we demonstrated how Alice can transmit the initial state of her detector coherently to the quantum field. The information about Alice's initial state is imprinted into the field observables in the two spacetime patches in which Alice couples to the field. (These are determined by the support of the detector profile function.) These observables propagate to the right at the speed of light since Alice coupled to the right-moving momentum of the field.

\begin{figure}
\centering
\includegraphics[width=0.45\textwidth]{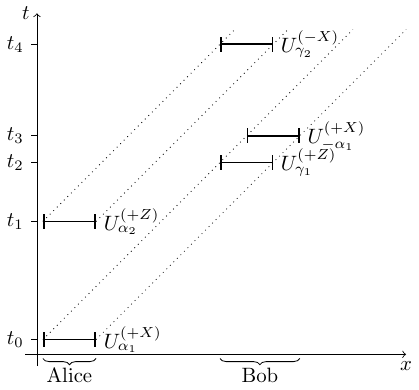}
  \caption{Spacetime diagram of the state transfer protocol. The dotted lines represent lightrays emanating from Alice's interaction with the field. The couplings correspond to controlled displacement operators as defined in \eqref{eq:ctrldisplop},\eqref{eq:U2},\eqref{eq:V1},\eqref{eq:V2} and \eqref{eq:V3}.  Displacements denoted by $\alpha_i$ correspond to a strong interaction with the field, whereas $\gamma_i$ denote weak interactions with the field which Bob uses to sense Alice's displacement of the field.}
  \label{fig:spacetimediag} 
\end{figure}

In order to achieve quantum state transfer from Alice's to Bob's detector, Bob has to retrieve the information about Alice's initial state from the field.
We will now show how to do this using a sequence of three interactions between Bob's detector and the field (see the spacetime diagram in Figure \ref{fig:spacetimediag}) that essentially implement a SWAP gate. 
Our protocol makes use of the methods developed in \cite{munro_entangled_2000,leghtas_deterministic_2013}.

The first and the third coupling are designed to change the state of Bob's detector conditional on the field state. For the fidelity of the state transfer protocol {\blu to be high,} it is important that these interactions change the field state as little as possible.
The second interaction between Bob and the field is designed to erase information about Alice's initial state from the field. 
More specifically, by acting on the field conditional on Bob's state, 
it undoes the displacement that Alice imprinted on the field in her first interaction, which was specified by $\alpha_1$.
All three interactions have the same structure as Alice's interactions, and consequently the corresponding unitaries 
also have the form of controlled displacement operators, as in \eqref{eq:ctrldisplop}. 

Bob's  coupling parameters need to be chosen so as to reflect the different role of the interactions and, of course, need to be tuned to Alice's choice of coupling parameters $\alpha_1$ and $\alpha_2$. 
However, we note that  Alice's and Bob's coupling parameters are independent of which particular state is being transmitted, 
and therefore our protocol does not require a classical side channel.

In the following we discuss how the three couplings transfer Alice's message from the field to Bob's detector. 
We recall that an ideal state transfer protocol would be achieved if Bob's detector ended up in the pure state $\ket\psi_B=x_+\ket{+X}_B+x_-\ket{-X}_B$. However, the scheme presented here is only an approximate state transfer protocol. Nevertheless, Bob's final state can be brought arbitrarily close to the target state.
We discuss below how Bob's couplings need to be designed to achieve this goal. 
Appendix \ref{app:overlap} gives a detailed calculation of the overlap between Bob's final state and the ideal target state.

\subsection{Bob sensing the displacement of the field}
In full generality, Alice's interaction with the field left her detector and the field 
in the state \eqref{eq:stateafterAlice}.
In order to simplify the following discussion, we will explicitly consider the cases where Alice sent one of the basis states $\ket{\pm X}_A$, and subsequently combine the two cases to address arbitrary states.

If Alice's detector started in the state $\ket{+X}_A$, then the field and her detector are now in the state
\begin{align}
    \ket{{x_+}}_{A,F} &= U\ket{+X}_A\ket0_F\nn* 
    &=\frac1{\sqrt2} \left( \ket{+Z}_A  \ket{\alpha_2+\alpha_1}   +\ket{-Z}_A \ket{\alpha_1} \right),
\end{align}
whereas $\ket{-X}_A$ evolved to
\begin{align} 
    \ket{x_-}_{A,F} &= \frac1{\sqrt2} \left( \ket{+Z}_A  \ket{\alpha_2} -\ket{-Z}_A \ket0 \right).
\end{align}
Note  
that the field is displaced by $\alpha_1$ if Alice started in the state $\ket{+X}_A$, but not if she started in $\ket{-X}_A$.  (This displacement by $\alpha_1$ is in addition to a possible displacement by $\alpha_2$, which we will address later on.) The first coupling is designed to transfer this information onto Bob's detector.

We assume that Bob couples his detector to the field exactly at a time $t=t_2$ when the lightrays emanating from Alice's first coupling reach him. (See Figure \ref{fig:spacetimediag}.) 
Furthermore, we take Bob's detector to be initialized in the state $\ket{-X}_B$ and choose the first coupling such that it flips Bob's detector state to $\ket{+X}_B$ if the field is displaced along $\alpha_1$, but leaves it unchanged otherwise. 
That is, we want $V^{(1)}$ to yield  
\begin{align}\label{eq:desierablesV1}
 &V^{(1)}\ket{x_+}_{A,F}\ket{-X}_B\nn* 
 &\approx \frac{-1}{\sqrt2} \ket{+X}_B \left( \ket{+Z}_A   \ket{\alpha_2+\alpha_1} + \ket{-Z}_A \ket{\alpha_1} \right)
 \end{align}
while
\begin{align} \label{eq:desierablesV2}
 &V^{(1)}\ket{x_-}_{A,F}\ket{-X}_B\nn* 
 &\approx \frac1{\sqrt2}\ket{-X}_B \left( \ket{+Z}_A  \ket{\alpha_2} -\ket{-Z}_A \ket0\right)
\end{align}
up to some small error term.
This can be achieved by a unitary of the same form as in previous steps,
\begin{align}\label{eq:V1}
V^{(1)}&= U_{\gamma_1}^{(+Z)}\nn
&=\ketbra{+Z}{+Z}_B \otimes D_{\gamma_1}+\ketbra{-Z}{-Z}_B\otimes\id\nn 
&=\frac12\left[ \left(\ketbra{+X}{+X}+\ketbra{-X}{-X}\right) \otimes (D_{\gamma_1}+\id) \right.\nn*
&\qquad \left. + \left(\ketbra{+X}{-X}+\ketbra{-X}{+X}\right)\otimes (D_{\gamma_1}-\id) \right],
\end{align}
if  the displacement $\gamma_1$ is chosen such that 
\begin{align}\label{eq:desirablesDg1a}
&D_{\gamma_1}\ket{0}\approx+\ket{0}, ~D_{\gamma_1}\ket{\alpha_2}\approx+\ket{\alpha_2}
\end{align}
while 
\begin{align}\label{eq:desirablesDg1b}
&D_{\gamma_1}\ket{\alpha_1}\approx-\ket{\alpha_1} , ~D_{\gamma_1}\ket{\alpha_2+\alpha_1}\approx-\ket{\alpha_2+\alpha_1}.
\end{align}
In the following we discuss how such a displacement operator can be constructed.

First, note that the action of $D_{\gamma_1}$  is generally independent of whether the field is displaced by $\alpha_2$ or not, because the two couplings $V^{(1)}$ and $U^{(2)}$ are spacelike separated (see Figure \ref{fig:spacetimediag}), and therefore the respective displacement operators commute, $\comm{D_{\gamma_1}}{D_{\alpha_2}}=0$ (see \eqref{eq:phifromcomm}).

In order for $D_{\gamma_1}$ to have the desired effect, the displacement $\gamma_1$ has to fulfill two requirements, namely 
\begin{align}
\label{eq:gammacondition1}
\varphi(\gamma_1,\alpha_1):=\Im\integral{k}{}{} \gamma_1(k) \alpha_1(k)^*=\frac\pi2
\end{align}
and
\begin{align}
\label{eq:gammacondition2}
\|\gamma_1\|^2:=\integral{k}{}{}|\gamma_1(k)|^2<<1 .
\end{align}
Condition  \eqref{eq:gammacondition1} ensures that, under the action of $D_{\gamma_1}$, the states that are displaced by $\alpha_1$ acquire a phase, $\ee{\ii\varphi(\gamma_1,\alpha_1)}=\ee{\ii\pi/2}$, in addition to the displacement by $\gamma_1$. That is,
\begin{align}
&D_{\gamma_1}\ket{\alpha_1}= \ee{\ii\varphi(\gamma_1,\alpha_1)}\ket{\gamma_1+\alpha_1}\\  
&D_{\gamma_1}\ket{\alpha_1+\alpha_2}=  \ee{\ii\varphi(\gamma_1,\alpha_1)}\ket{\gamma_1+\alpha_1+\alpha_2}
\end{align}
while
\begin{align}
&D_{\gamma_1}\ket{0}= \ket{\gamma_1},  
&D_{\gamma_1}\ket{\alpha_2}=\ket{\alpha_2+\gamma_1},
\end{align}
as follows from the composition formula for displacement operators \eqref{eq:displacementcomposition}.

These states differ from the desired outcomes by an additional displacement $\gamma_1$. Now condition \eqref{eq:gammacondition2} ensures that this displacement
is small, so that the error in \eqref{eq:desierablesV1} and \eqref{eq:desierablesV2} is also small.
In fact, by \eqref{eq:coherentoverlapepsilon}, we have, e.g.,
\begin{align}
&\ket{\gamma_1+\alpha_1}\sim \ee{\ii\varphi(\gamma_1,\alpha_1)} \ket{\alpha_1} +\mOgamma 
\end{align}
such that
\begin{align}\label{eq:DonAlpha}
D_{\gamma_1}\ket{\alpha_1}&\sim -\ket{\alpha_1}+\mOgamma
\end{align}
and analogously $D_{\gamma_1}\ket{\alpha_1+\alpha_2}\sim -\ket{\alpha_1+\alpha_2}+\mOgamma$, which is what we required in \eqref{eq:desirablesDg1b}.
Therefore, the interaction $V^{(1)}$ as defined in \eqref{eq:V1} fulfills equations \eqref{eq:desierablesV1} and \eqref{eq:desierablesV2} up to error terms of order $\mathcal{O}\left(\|\gamma_1\|^2\right)$, given that $\gamma_1$ fulfills the two requirements \eqref{eq:gammacondition1} and \eqref{eq:gammacondition2}.

The question now is how Bob can design a coupling that fulfills requirements \eqref{eq:gammacondition1} and \eqref{eq:gammacondition2}. Since the signal is encoded into the momentum of the field, it may appear natural to have Bob read out the signal by coupling to the amplitude of the field. This is possible, as we discuss in Appendix \ref{app:amplitudecoupling}, but in the present (1+1)-dimensional setting, this gives rise to well-known infrared divergences which need to be addressed carefully.

The problem of IR divergence can be avoided altogether by coupling Bob to the right-moving momentum of the field, like Alice. Of course, in order for Bob to be able to read out any information, he must couple to the field through a field observable that does not commute with the observable through which Alice coupled 
(see \eqref{eq:Hint1}). 
This can be achieved if Bob uses a different detector profile from Alice. For example, Alice could use an asymmetric profile function and Bob the mirrored version thereof.

Based on these considerations, we let Bob couple to the right-moving field momentum, with
\begin{align}
\intHd{B}^{(1)}=\frac{\mu_B}2 \delta(t-t_2) \left(\id-\sigma_X\right) \otimes\integral{x}{}{} g(x)\pi_-(x,t),
\end{align}
to generate $V^{(1)}$. Then
\begin{align}
&\varphi(\gamma_1,\alpha_1)\nn* 
&= -\frac\ii2 \comm{\mu_A \integral{x}{}{} f(x) \pi_-(x,t_0) }{\mu_B \integral{x}{}{} g(x) \pi_-(x,t_2) }\nn 
&=\frac14\mu_A\mu_B \integral{x}{}{}  f'\left(x\right) h(x),
\end{align}
where $h(x)=g(x+(t_2-t_0))$ is Bob's shifted profile function, and we use the commutation relation (see, e.g., \cite{jonsson_information_2016})
\begin{align}
{\comm{\pi_-(x,t_0)}{\pi_-(y,t_2)}}={\frac{-\ii}{2} \delta'( (t_2-t_0)-(y-x))}.
\end{align}
Here, $\delta'(x)$ is the distributional derivative of the Dirac $\delta$-distribution, satisfying $\integral{x}{}{} \delta'(x)f(x)=-f'(0)$.

This allows us to fulfill both requirements on $\gamma_1$, \eqref{eq:gammacondition1} and \eqref{eq:gammacondition2}: choosing
\begin{align}
\frac1{\mu_B}=\frac{\mu_A}{2\pi} \integral{x}{}{}  f'\left(x\right) h(x),
\end{align}
ensures that $\varphi(\gamma_1,\alpha_1)=\pi/2$, while
\begin{align}
\gamma_1(k)= \mu_B\ee{\ii k t_2} \sqrt{\frac{k}{4\pi}}  \tilde{g}(k) \theta(k)
\end{align}
makes $\|\gamma_1\|^2$   finite and inversely proportional to Alice's coupling strength $\mu_A$.

\subsection{Bob acting back on the field}\label{sec:secondBobcoupling}

The objective of the second coupling is to undo the displacement of the field along $\alpha_1$, so as to delete this piece of information from the field. Since Bob has just read out whether the field is displaced along $\alpha_1$, he can use the coupling 
\begin{align}\label{eq:V2}
V^{(2)}&=U_{-\alpha_1}^{(+X)}\nn
&=\ketbra{+X}{+X}_B \otimes D(-\alpha_1)+\ketbra{-X}{-X}_B\otimes\id,
\end{align}
which is essentially the inverse of Alice's first coupling. If Alice, in her first coupling, displaced the field by $\alpha_1$, then Bob's detector, after his first coupling, is now in the state $\ket{+X}_B$. In this case, his second coupling undoes the displacement by $D(-\alpha_1)$: \begin{align}\label{eq:stateAfaftertwox+}
V^{(2)} V^{(1)}\ket{x_+}&\sim \frac{-1}{\sqrt2} \ket{+X}_B \left( \ket{+Z}_A   \ket{\alpha_2} + \ket{-Z}_A \ket{0} \right) \nn* 
&\qquad+\mOgamma,
\end{align}
where we abbreviate $\ket{x_+}=\ket{x_+}_{A,F}\ket{-X}_B$. If, on the other hand, Bob remained in the state $\ket{-X}_B$ after his first coupling, then the field state is also unchanged by $V^{(2)}$:
\begin{align}
V^{(2)} V^{(1)}\ket{x_-}&\sim \frac1{\sqrt2}\ket{-X}_B \left( \ket{+Z}_A  \ket{\alpha_2} -\ket{-Z}_A \ket0\right)\nn* 
&\qquad+\mOgamma. \label{eq:stateAfaftertwox-}
\end{align}

A challenging feature of this second coupling  is that, in order to realize the displacement operator $D_{-\alpha_1}$, Bob  needs to interact with the same field observables as Alice did in her first interaction. However, these are  the same observables with which Bob already had to interact in his first coupling. This raises the question of how Bob can access the field observables, which are propagating at the speed of light, at two different points in time.

A simple answer to this question is to put the entire setup in a Dirichlet cavity: Here Bob can just wait for Alice's signal to return to him after it is reflected by the cavity walls, and implement $V^{(2)}$ then. (Since Bob's second and third coupling, which will be defined in \eqref{eq:V3}, always commute by construction, Bob can swap the order of the second and third couplings if necessary.)
In free Minkowski spacetime the situation is more difficult: in this setting, in order to access the observables to which Alice coupled at two different points in time, Bob's detector profile would have to shift along with the signal's propagation, at the speed of light.

Furthermore, we note that a model that assigns finite spatial extension to the detector fully respects causality if predictions are restricted to timescales longer than the light-crossing time of the detector's spatial extension \cite{martin-martinez_causality_2015}. On shorter scales, the assumption of spatially extended detectors leads to the problematic implication that the interaction between the detector and the field takes places simultaneously at space-like separated points. For example, in the protocol at hand, 
in order for the interaction of Bob and the field on the left boundary of Bob's detector profile during the first interaction $V^{(1)}$ to have an influence on the interaction on the right boundary of Bob's detector profile during the second interaction $V^{(2)}$, the information obtained in the first interaction would have to propagate faster than light across the detector.

\subsection{Bob disentangling from Alice and the field}

The first two couplings of Bob's detector are sufficient to bring it into the correct final state if Alice's initial state was either $\ket{+X}_A$ or $\ket{-X}_A$. However, for general initial states, i.e., superpositions of these two states, state transfer is not complete yet. This is because Bob, in general, is still entangled with the field and Alice after his first two couplings, so that his partial state is an incoherent mixture of $\ket{+X}_B$ and $\ket{-X}_B$. In order to put Bob's detector in the final (pure) target state, the third coupling needs to disentangle Bob from the field and Alice.

We see in equations \eqref{eq:stateAfaftertwox+} and \eqref{eq:stateAfaftertwox-} that Bob's entanglement with Alice and the field arises because the two states of  Alice and the field, $\frac1{\sqrt2}\left(\mp\ket{+Z}_A   \ket{\alpha_2} - \ket{-Z}_A \ket{0}\right) $, are orthogonal to each other.
This can be remedied by choosing Bob's third coupling such that it applies a phase $-1$ if Bob is in the state ${-X}_B$ and the field is displaced by $\alpha_2$. 
That is, let
\begin{align}\label{eq:V3}
V^{(3)}&=U_{\gamma_2}^{(-X)}\nn
&=\ketbra{-X}{-X}_B \otimes D_{\gamma_2}+\ketbra{+X}{+X}_B\otimes\id,
\end{align}
where $\gamma_2$ is chosen such that 
\begin{align}\label{eq:requirementdelta}
\varphi(\gamma_2,\alpha_2)=\frac\pi2\text{ and }\|\gamma_2\|^2<<1.
\end{align}
This coupling $V^{(3)}$ relates to Alice's second coupling in the same way as Bob's first coupling relates to Alice's first coupling.
As indicated in Figure \ref{fig:spacetimediag}, it takes place at time $t=t_4$ when the lightrays from Alice's second coupling reach Bob. 

For simplicity, we assume that both of Alice's couplings use the same profile function of the detector, so that $\|\alpha_1\|^2=\|\alpha_2\|^2$, and accordingly $\|\gamma_1\|^2=\|\gamma_2\|^2$. This allows us to express the errors in the subsequent calculations in terms of a single parameter.

Analogously to \eqref{eq:DonAlpha}, we have $D_{\gamma_2}\ket{\alpha_2}\sim-\ket{\alpha_2}+\mOgamma$, therefore
\begin{align}
&V\ket{x_-}= V^{(3)}V^{(2)} V^{(1)}\ket{x_-}\nn* 
&\sim \frac{-1}{\sqrt2}\ket{-X}_B \left( \ket{+Z}_A  \ket{\alpha_2} +\ket{-Z}_A \ket0\right)+\mOgamma,
\end{align}
whereas if Alice started in $\ket{+X}_A$, then Bob's final coupling has no effect,
\begin{align}
&V\ket{x_+}\sim V^{(2)} V^{(1)}\ket{x_+}\mOgamma\nn* 
&\sim\frac{-1}{\sqrt2} \ket{+X}_B \left( \ket{+Z}_A   \ket{\alpha_2} + \ket{-Z}_A \ket{0} \right)+\mOgamma.
\end{align}
We see that after Bob's final coupling, the field and Alice are in the same state in both cases.
This means that also for an arbitrary  state $\ket\psi=x_+\ket{+X}+x_-\ket{-X}$ we obtain
\begin{align}
&VU \ket\psi_A \ket0\ket{-X}_B \nn* 
&\sim \frac1{\sqrt2} \left( \ket{+Z}_A   \ket{\alpha_2} + \ket{-Z}_A \ket{0} \right) \ket\psi_B+\mOgamma.
\end{align}
In other words, up to order $\mOgamma$, Bob's detector ends up in the pure state that Alice initially sent. In fact, as shown in Appendix  \ref{app:overlap}, the overlap of Bob's exact final state $\rho_B$ and the ideal (pure) target state is lower-bounded by
\begin{align}
&\tr\left(  \rho_B \ketbra\psi\psi_B\right) \geq 1-\frac12 \|\gamma_1\|^2.
\end{align}
We conclude that arbitrarily state transfer is possible if Bob's sensing interactions can be designed such that the field states are hardly displaced by them, i.e., $\|\gamma_1\|^2,\|\gamma_2\|^2<<1$ are very small.

\section{Delocalizing quantum information in a cavity field}\label{sec:delocalizingcavity}

The previous section showed that quantum state transfer is possible between detectors that couple to the right-moving momentum of the field. In this way, the complete information about Alice's initial state propagates towards a single receiver without being dispersed in different directions. Consequently, a single Bob is able to receive all the information and recover Alice's initial state from the signal.

Coupling Alice to the field symmetrically  would be a hindrance to the transmission of quantum information: If Alice couples symmetrically to both the left- and right-moving observables, emitting equally in both directions, then a receiver must have access to both parts of the signal in order to retrieve Alice's initial state from the field.

However, this particular obstacle to quantum state transfer may be a key feature for implementing other information processing tasks, akin to quantum bit commitment or quantum secret sharing \cite{kent_unconditionally_1999,hillery_quantum_1999,gottesman_theory_2000,adlam_device-independent_2015}, because it forces receivers to cooperate if they want to retrieve quantum information from the sender's signal.

As a first step towards such implementations, we here consider a scenario  related to quantum state merging \cite{horodecki_quantum_2006}: The sender encodes a qubit state  into the relativistic field, in such a way that it is delocalized between two parts of the signal, propagating in opposite directions. Since the signals propagate at the speed of light, they cannot  be accessed by a single localized observer, but only by two parties who cooperate.
For example, one party could reflect the signal with a mirror, or else both parties capture their respective parts of the signal, then bring their detectors together and perform a joint unitary on them.
However, from the point in time when the sender emits the state until the point in time when the two parts of her signal can first be reunited, the message is delocalized in the  field. The field's relativistic properties ensure that it is inaccessible to any (localized) party in the mean time.\footnote{We recall that, as discussed in the introduction, we preclude classical communication via additional side-channels between the parties.}

One possible way to implement such a protocol is  based on the same couplings discussed in the previous sections: 
Alice performs essentially the same steps as in Section \ref{sec:AliceCoupling}, but now coupling to the full conjugate momentum of the field, $\pi=\pi_-+\pi_+$, including both left- and right-moving momentum.
Two receivers, one on the left $(L)$ and one on the right $(R)$ of Alice, can then extract Alice's initial state from the field by coupling their detectors to the left-moving (respectively right-moving) momentum of the field, as in Section \ref{sec:BobCoupling}. One additional modification is necessary in the final step, when the Bobs seek to get disentangled from Alice and the field: here, each Bob must acquire only half of the phase compared to the original protocol \eqref{eq:requirementdelta}. That is, they must choose $\gamma_2$ such that $\varphi(\gamma_2,\alpha_2)=\frac\pi4$. 
This ensures that an arbitrary initial state $\ket\psi=x_+\ket{+X}+x_-\ket{-X}$ of Alice leads to the final state
\begin{align}\label{eq:tworeceiverfinalstate}
 &\ket\psi_A\ket{-X}_L\ket{-X}_R\ket0 \nn 
 &\quad\mapsto \frac1{\sqrt2} \left( \ket{+Z}_A   \ket{\alpha_2} + \ket{-Z}_A \ket{0} \right) \otimes\nn* 
 &\qquad\qquad \left(x_+ \ket{+X}_L\ket{+X}_R + x_-\ket{-X}_L\ket{-X}_R\right),
\end{align}
where Alice's initial state is now encoded in a generally entangled joint state of the two receivers.

\begin{figure}
\centering
\includegraphics[width=0.45\textwidth]{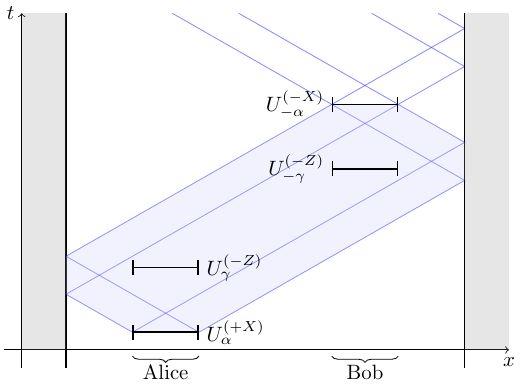}
  \caption{Spacetime diagram of the protocol discussed in Section \ref{sec:delocalizingcavity}  which delocalizes Alice's initial detector state into left- and right-moving modes of the cavity. Since Alice emits her initial state equally to the right and left it can only be retrieved by Bob at the specific focal points where the two parts of the signals periodically recombine. In between those points the qubit state is inaccessible. }
  \label{fig:cavity} 
\end{figure}

In the following we discuss a slightly different scenario, depicted in Figure \ref{fig:cavity}, which couples the detectors to the amplitude of a (massless, scalar Klein-Gordon) field inside a Dirichlet cavity. 
This comes with several advantages:

Inside a cavity, the protocol only requires a single receiver, because Alice's signal is reflected by the cavity walls such that both parts of the signal naturally recombine periodically.
Therefore, as far as potential future experimental implementations are concerned, a Dirichlet cavity may not only resemble an experimental setup more closely, but also allow for a less complex implementation.

A second advantage is that realizing  the protocol inside a cavity reduces the total number of couplings between detectors and field to four (from five in the previous section) and also avoids the causality issue discussed in Section  \ref{sec:secondBobcoupling}.
This is possible because inside the cavity the detectors can be coupled to the field \emph{amplitude} rather than the field momentum.
In  free 1+1 dimensional Minkowski spacetime, this would have been problematic because of IR divergences, as discussed in Appendix \ref{app:amplitudecoupling}.
Inside  a Dirichlet cavity, on the other hand, the discrete mode structure of the field and the absence of a zero-mode ensure that there occur no infrared divergences.

This allows us to make use of the particular property of the field amplitude commutator, which  is constant between strictly timelike separated points in 1+1 dimensions \cite{jonsson_information_2016}. This  we use  to implement two non-commuting couplings between the field and a detector without having to move the detector at all.

\subsection{Disentangling Alice from the field}

Using these features of the cavity setting,
we choose Alice's couplings to the field such that they already disentangle her detector from the field. This is in contrast with the previous protocol, which left Alice's detector and the field maximally entangled. Disentangling Alice from the field is possible by designing Alice's second coupling in a manner analogous to Bob's first coupling to the field.

Just as in Section \ref{sec:AliceCoupling}, we denote Alice's initial state by $\ket\psi=x_+\ket{+X}+x_-\ket{-X}$ and assume that the field starts out in the vacuum $\ket0$. Formally, also the first coupling between Alice and the field looks identical
\begin{align}
 U^{(1)}&= U^{(+X)}_{\alpha}=\ketbra{+X}{+X}_A\otimes D_\alpha+\ketbra{-X}{-X}_A\otimes\id .
\end{align}
However, there are a few differences between this coupling and the first coupling of Section \ref{sec:AliceCoupling}: Most importantly, the interaction here is generated by coupling the detector to the field amplitude, instead of the field momentum, through
\begin{align}\label{eq:secVAlicesfirst}
 \intH^{(1)}&= \lambda_1 \integral{x}{}{} f(x) \phi(x,t_0)\otimes\ketbra{+X}{+X}.
\end{align}
Note that for the coupling   to the field amplitude, the coupling constant $\lambda_1$ is dimensionless, in contrast to the coupling constants for the field momentum in the previous sections.

Inside the cavity the field  is expanded into the discrete set of modes
\begin{align}
 \phi(t,x)&= \sum_{j=1}^\infty \frac{1}{\sqrt{j \pi} } \sin(j \pi x/L) \left( a_j \ee{-\ii \frac{ j \pi }{L} t } + a^\dagger_j \ee{\ii \frac{j \pi }{L}t} \right).
\end{align}
Accordingly, the multi-mode displacement operator $D_{\alpha}$ now acts on a discrete set of modes, instead of a continuous set of modes. 
This means that the displacement amplitude is now a function of the mode number $j$. In particular, Alice's first coupling \eqref{eq:secVAlicesfirst} results in a displacement
\begin{align}\label{eq:discretealpha}
 \alpha(j) &= \frac{-\ii \ee{\ii\pi t_0/L}}{\sqrt{j\pi}}\tilde{f}_j, \qquad j=1,2,...
\end{align}
with $\tilde{f}_j= \integral{x}0L f(x)\sin(j\pi x/L)$.
Accordingly, for discrete modes we define
\begin{align}
 &\|\alpha\|^2=\sum_j \|\alpha(j)\|^2,\quad \varphi(\alpha,\epsilon) =\Im\sum_j \alpha(j)\epsilon(j)^*.
\end{align}
Adapting the  formulae in Appendix \ref{app:coherent}  amounts to replacing the momentum space integrations $\integral{k}{}{}\to\sum_{j=1,2,...}$ by sums over the mode number. 

After the first coupling, Alice and the field consequently are in the state
\begin{align}
 U^{(1)} \ket\psi\ket0&= x_+\ket{+X}\ket{\alpha}+x_-\ket{-X}\ket0.
\end{align}
which formally is the same as \eqref{eq:secIIIafterU1}, but with the displacement now given by \eqref{eq:discretealpha}.

The second coupling between Alice and the field is different from the second coupling in Section \ref{sec:AliceCoupling}. Instead of acting strongly for a second time on the field, we now use a  sensing interaction, like Bob's first coupling. It flips the detector state conditioned on the field state, but its effect on the field, in turn, is negligible. We  denote it by
\begin{align}
U^{(2)} &= U^{(-Z)}_{\gamma}= \ketbra{-Z}{-Z}_A\otimes D_\gamma+\ketbra{+Z}{+Z}_A\otimes\id
\end{align}
and require $\varphi(\gamma,\alpha)=-\pi/2$ and $\|\gamma\|^2<<1$. From these requirements, as shown in \eqref{eq:DonAlpha}, it follows that $D_\gamma \ket{\alpha}\sim -\ket{\alpha}+\mathcal{O}(\|\gamma\|^2)$, whereas $D_\gamma\ket0\sim \ket0+\mathcal{O}( \| \gamma\|^2)$. Hence, 
\begin{align}\label{eq:disentangledAlicestate}
 U^{(2)}U^{(1)} \ket\psi\ket0&\sim \ket{-X}\left(x_+\ket{\alpha}+x_-\ket0\right) +\mathcal{O}( \| \gamma\|^2)
\end{align}
which means that, up to corrections of order $\mathcal{O}( \| \gamma\|^2)$, Alice and the field are left in a product state. In contrast to this, note that in Section  \ref{sec:AliceCoupling} Alice and the field ended up in an (almost) maximally entangled state. (See  \eqref{eq:stateafterAlice} and discussion thereafter.)

To implement the second  coupling we can use the fact that the commutator of the field amplitude of a massless field has timelike support:
In free Minkowski spacetime it is constantly $\comm{\phi(x,t_0)}{\phi(y,t_1)}=\ii/2$ if $(y,t_1)$ is inside the future lightcone of $(x,t_0)$. Inside a Dirichlet cavity this still holds true if by the time $t_1$ no lightrays reflected by the cavity walls have reached from $x$ to $y$ (see, e.g., \cite{jonsson_information_2016}).
This allows Alice to implement the second coupling inside the future lightcone of her first coupling. 
Thus we  avoid the causality issue discussed in Section \ref{sec:secondBobcoupling}, where Bob had to move at the speed of light in order to couple to the same field observables twice.

We choose  Alice's second coupling to take place at time $t=t_1$ such that the delay $t_1-t_0$ is long enough for the couplings to be timelike separated, but short enough such that reflected lightrays emanating from the first coupling have not yet returned to Alice.
Then, if we denote the interaction Hamiltonian of the second interaction by
\begin{align}
    \intH^{(2)} = \lambda_2 \integral{x}{}{} f(x) \phi(x,t_1)\otimes\ketbra{-Z}{-Z},
\end{align}
we obtain, by \eqref{eq:phifromcomm},
\begin{align}
\varphi(\gamma,\alpha)  = -\frac{\lambda_1 \lambda_2}{4} \left(\integral{x}{}{}f(x)\right)^2.
\end{align}
Assuming a normalized profile function, $\integral{x}{}{}f(x)=1$, we can achieve the first requirement on $\gamma$ by choosing the coupling constant of the second coupling to be
\begin{align}
 \lambda_2=\frac{2\pi}{\lambda_1}.
\end{align}
The coupling constants being inversely proportional to each other also helps to fullfill the second requirement of $\|\gamma\|^2<<1$, because
\begin{align}
\|\gamma\|^2= \frac{4\pi^2}{(\lambda_1)^4}  \|\alpha\|^2 \sim \frac1{\left(\lambda_1 \right)^2 }
\end{align}
as $\lambda_1\to\infty$ increases.

\subsection{Bob reading out the field state}

The advantage of disentangling Alice from the field is that it allows Bob to read  out the transmitted qubit state with just two couplings, instead of the three couplings of Section \ref{sec:BobCoupling}. 

Since Bob starts in the state $\ket{-X}$, it is easy to see from \eqref{eq:disentangledAlicestate} that the couplings that allow Bob to transfer the state from the field into his own detector are given by unitaries which are exactly the inverses of the $U^{(2)}$ and $U^{(1)}$: this gives
\begin{align}
& V^{(2)}V^{(1)} U^{(2)}U^{(1)} \ket\psi\ket0\ket{-X}_B \nn* 
 &\qquad\sim \ket{-X}_A \ket0 \ket\psi_B+\mOgamma.
\end{align}

More precisely, Bob's first coupling needs to correspond to the inverse of Alice's second coupling. This means it should read
\begin{align}
V^{(1)} &= U^{(-Z)}_{-\gamma} =\ketbra{-Z}{-Z}\otimes D_{-\gamma}+\ketbra{+Z}{+Z}\otimes\id, 
\end{align}
such that 
\begin{align}
& V^{(1)} \ket0\ket{-X}_B  \nn* 
&\qquad \sim \ket{-X} x_-\ket0 + \ket{+X} x_+\ket{\alpha_j} +\mathcal{O}( \| \gamma\|^2).
\end{align}
(The same can also be achieved with $V'^{(1)}=  \ketbra{-Z}{-Z}\otimes D_\gamma+\ketbra{+Z}{+Z}\otimes\id$, since 
the relevant phase factor is $\ee{\ii2\varphi(\gamma,\alpha)}=\ee{\ii2\varphi(-\gamma,\alpha)}=-1$ in either case.)

Just as Alice's second coupling, this first coupling of Bob can be implemented inside the lightcone of Alice's first coupling, before the reflected lightrays from Alice's first coupling reach Bob's location.
Bob's second coupling needs to be timed more carefully, and it requires Bob to be located exactly where the lightrays from Alice's first interaction intersect again, after being reflected by the opposite cavity walls. This is because Bob needs to undo the field displacement from Alice's first interaction with his second coupling,
\begin{align}
 V^{(2)}&= U^{(+X)}_{-\alpha}=\ketbra{+X}{+X}_A\otimes D_{-\alpha}+\ketbra{-X}{-X}_A\otimes\id.
\end{align}

These restrictions on the location for retrieving Alice's initial state ensure that, once Alice has injected her message into the field, it cannot be coherently extracted again before at least one light cavity crossing time has passed.

\section{Conclusions and outlook}

Studying the concrete case of a relativistic quantum field in 1+1 spacetime dimensions, we have given a non-perturbative account of wireless quantum communication between localized observers, which were modeled as Unruh-DeWitt particle detectors. 
In particular, going beyond previous literature, we have developed a protocol which performs approximate quantum state transfer   between particle detectors and can in principle come arbitrarily close to optimal quantum capacity.
The  1+1 dimensional scenario is particularly interesting since it allows one to explore novel methods of quantum information processig that may be implementable, e.g., in superconducting circuits.

To build the quantum state transfer protocol, we extended coherent state methods developed for the transfer of qubit states to a single harmonic mode   \cite{leghtas_deterministic_2013}. We extended this to the coupling between a localized model atom and the many modes (continuous or discrete) of a relativistic field.

An implementation of the presented protocols in higher dimensional spacetimes could be achieved by coupling the signalling devices to field observables with a narrow directional profile, similar to the ones considered in \cite{downes_quantum_2013,bruschi_spacetime_2014}.

For generalizations to higher spacetime dimensions it will also be important to consider the particular consequences of multi-directional signal emission for quantum communication. In classical wireless communication the multi-directionality  is a valuable feature and only causes a quantitative loss of signal strength.
However, in quantum wireless communication the no-cloning theorem presents a qualitatively different challenge to wireless communication.
In this context, in Appendix \ref{sec:symmetriccouplingissues} we give a rigorous account of how this problem manifests in a scenario with symmetric emission to several receivers, showing that the quantum channel from the sender to any single receiver is anti-degrabable and, thus, has zero quantum capacity.
On the other hand,  we review  how  symmetric signals could be specifically designed such that different receivers are required to cooperate in order to retrieve the quantum information emitted by the sender. This could be of interest for quantum information processing tasks similar to quantum secret sharing or quantum bit commitment \cite{kent_unconditionally_1999,hillery_quantum_1999,gottesman_theory_2000}.

In conclusion, aiming to advance our understanding of quantum fields from an information-theoretical point of view, we have proposed a prototype framework for the study of quantum information transmission between local signaling devices through  quantum fields.
Further, it should also be very interesting to pursue the fundamental implications by extending the present study to general relativistic settings such as expanding universe scenarios and, in particular, to the question of the extent to which black holes broadcast classical and quantum information in Hawking radiation.

\section*{Acknowledgements}
RHJ is grateful to G\"oran Johansson for helpful discussions and for bringing \cite{leghtas_deterministic_2013} to his attention. 
RHJ  acknowledges support from the Knut and Alice Wallenberg Foundation,
ERC Advanced grant 321029 and by the VILLUM FONDEN via the QMATH Center of Excellence (grant no.10059).
Research at Perimeter Institute is supported by the Government of Canada through Industry Canada and by the Province of Ontario through the Ministry of Research and Innovation. AK and EMM acknowledge support from the Discovery Program of the National Science and Engineering Research Council of Canada (NSERC).
\appendix

\section{Symmetry and Wireless Quantum Communication} \label{sec:symmetriccouplingissues}

In classical wireless communication antennae are often designed to emit their signals symmetrically, so as to be able to reach receivers in many different directions which all receive identical signals.
The loss of signal power that results from distributing the signal can be compensated for by the receivers, e.g., by the use of amplifiers.
This approach is impossible in quantum wireless communication because quantum information cannot be cloned. Here the symmetric emission of wireless signals poses a fundamental obstacle to quantum information transmission. 

Intuitively, one can see directly from the no-cloning principle that the channel from the sender to any single such receiver must have zero quantum capacity. 
In this appendix,
we give a precise formulation of this argument, and review the required notions from quantum information theory along the way. 
We then discuss certain tasks of quantum information processing which are possible either despite, or because of the symmetric propagation of signals.

In order to keep the discussion general, we make no specific assumptions on the type of quantum signaling device being used. Instead, we only discuss general properties of the quantum channel between sender (Alice) and receiver (Bob), i.e., the map from Alice's input state to Bob's output state, or we use simple toy models for illustration.

\subsection{Symmetric emission results in vanishing quantum capacity}\label{sec:symmzeroqc}

\begin{figure}
\centering
\includegraphics[width=0.3\textwidth]{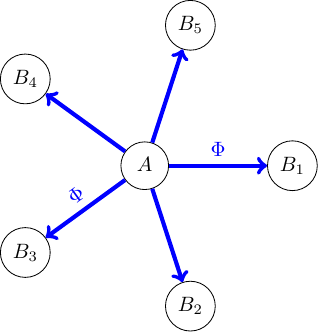}
  \caption{In a scenario where two or more receivers receive equal signals, the quantum capacity of the quantum channel $\Phi$ from the sender to any single such receiver vanishes. This is because $\Phi$ is an anti-degradable channel.}
  \label{fig:symmchannel} 
\end{figure}

When a sender emits signals symmetrically to a number of receivers -- that is, such that the resulting state is invariant under permutations of the receivers --, then the quantum capacity from the sender to any single receiver is zero, as illustrated in Fig. \ref{fig:symmchannel}.
This can be understood intuitively as a consequence of the no-cloning theorem:
Assume that two receivers receive equal signals from a sender and that this signal contains enough information to reconstruct the sender's initial state. Then both receivers could independently produce copies of that state, which is, of course, a violation of the  no-cloning theorem.

The remainder of this section provides a more formal argument to this end, based on the observation that the quantum channel from the sender to a single receiver in this scenario is anti-degradable, which in turn implies that its quantum capacity is zero. 
For the purpose of this discussion, we first review the notions of quantum capacity and anti-degradability.

\emph{Quantum capacity} measures a quantum channel's usefulness for transmitting quantum information; more concretely, for sharing entanglement: 
Suppose that Alice initially shares a generic entangled state, e.g., \mbox{$\ket{\psi}=\frac1{\sqrt2}\left(\ket0_A\ket1_C-\ket1_A\ket0_C\right)$} with a third party, Charlie. Alice then attempts to send her half of the state to Bob by passing her half through a quantum channel $\xi$.
If Alice succeeds, so that in the end Bob shares the state $\ket{\psi}$ with Charlie, then Alice has transmitted quantum information to Bob. In this ideal case the quantum channel would have maximal quantum capacity.

The quantum channel capacity
measures the rate at which a non-ideal, noisy channel can be used to transmit quantum information. It is given by the number of qubits of information which can be transmitted faithfully (with arbitrarily small error) per channel use, when many replicas of the channel are used in parallel. (For a thorough treatment of the topic, e.g.,  see \cite{hayashi_quantum_2017a,caruso_quantum_2014,wilde2013quantum}. In this general context, see also \cite{pirandola_fundamental_2017,laurenza_general_2017}.)

\emph{Anti-degradability} of a quantum channel is defined in terms of the concept of the complementary channel. It is rooted in the Stinespring dilation of the channel: 
Any given quantum channel $\Phi$, acting on states in $\mathcal{H}_A$, can be represented as resulting from a unitary $U_{AE}$ that acts on $\mathcal{H}_{A}\otimes\mathcal{H}_E$, where $\mathcal{H}_E$ represents some environment, 
as
\begin{align}
\Phi(\rho) = \tr_E \left[ U_{AE} (\rho \otimes \ketbra0{0}_E) U^\dagger_{AE} \right].
\end{align}
The complementary channel $\bar\Phi$ is then defined as
\begin{align}
 \bar\Phi(\rho)=\tr_A \left[ U_{AE} (\rho \otimes \ketbra0{0}_E) U^{\dagger}_{AE} \right]
\end{align}
and it maps an input state $\rho$ on $\mathcal{H}_A$ to the partial state of the environment after the joint evolution.

The channel $\Phi$ is called \emph{anti-degradable} if there exists another quantum channel $\Phi^R$ such that $\Phi = \Phi^R \circ \bar\Phi$, i.e., if the channel $\Phi$ can be obtained by composing the channel $\Phi^R$ and the complementary channel $\bar\Phi$ \cite{devetak2005capacity,hayashi_quantum_2017a,caruso_quantum_2014,wilde2013quantum}.
Physically speaking, this means that full information about the output state of an anti-degradable channel is contained in the final state of the environment. One can now see, by a no-cloning type of argument, that the quantum capacity of anti-degradable channels is zero \cite{holevo2008entanglement, caruso2006degradability}.

With these definitions and concepts in hand, we can now formulate the following statement:

\emph{Statement:} Consider a scenario with a single sender, $A$, and two or more equal receivers, $B_i$. Assume that sender and receivers communicate by coupling to an intermediary quantum system $F$, for instance a quantum field. 
One can then see that 
the channel $\Phi$ from $A$ to any single receiver, say $B_1$, 
is anti-degradable.

\emph{Proof:}
A unitary dilation 
of $\Phi$ is given by the unitary time evolution operator of the total system, which is composed of the signaling devices of sender and receivers as well as the field. 
It comprises the  unitary interaction between $A$ and $F$, between $F$ and the $B_i$, as well as the free evolution of all components. 

The channel $\Phi$, from $A$ to $B_1$, is obtained by tracing out the field and all other signalling devices $B_{i\neq 1}$.  Conversely, the complementary channel $\bar\Phi$ is obtained by tracing out only $B_1$. 
However, due to the symmetry of the signal, the partial state of $B_1$ is identical to that of any of the other receivers, which now are considered part of the environment.  Consequently, by composing $\bar\Phi$ with the partial trace over $F$ and all but one other receiver, we again obtain the channel $\Phi$. It follows that $\Phi$ is anti-degradable, and we conclude that  
the permutationally invariant signals to two or more receivers lead to vanishing quantum capacity for the channel from the sender to any single receiver. $\blacksquare$

This applies, for example, to the simple picture of wireless quantum communication where one qubit of information is encoded into the initial state of a sender atom which then emits this quantum information as a photon into the electro-magnetic field. Because the interactions in nature generally are symmetric, e.g., spherically symmetric or  left-right symmetric, the quantum channel from the sender to any one receiver that does not have access to more than half of the emitted radiation is necessarily zero. Therefore, highly directional emission profiles are essential in order to achieve wireless communication with non-zero quantum capacity.

\subsection{Quantum information processing with symmetric signals}\label{sec:qipwsymmetry}

The previous section showed that the quantum capacity is in general zero when the sender emits their signal symmetrically in different directions. 
While this is obviously an obstacle for tasks like quantum state transfer,  
there are other interesting tasks for which quantum capacity is not the relevant figure of merit and which can be accomplished  
with symmetrically emitted signals. 

Here we present two instructive examples of such tasks. The first one is heralded stochastic state transfer, in which a quantum state is transferred to one randomly chosen receiver out of several. In the second example, we discuss how symmetric signalling can be used to delocalize quantum information among many receivers, thereby forcing them to cooperate in order to retrieve the original message.

In similar ways the challenge of symmetric signal emission may be turned into a feature for the future implementation of tasks similar to quantum bit commitment or quantum secret sharing \cite{kent_unconditionally_1999,hillery_quantum_1999,gottesman_theory_2000,adlam_device-independent_2015}.

\subsubsection{Heralded stochastic state transfer}

Our first wireless communication scenario is based on the following physical observation:
quantum information that is sent using symmetrically emitted signals can be recovered coherently by a single, localized receiver, \emph{if one uses an additional degree of freedom} to encode where the message is being sent, thereby enabling receivers to post-select on having received a message at their location.
For example, in the simple case of an atom emitting a single photon in a superposition of directions, one can use the polarization of the photon to encode the message, while the location of the photon heralds at which location the message is received.

A minimal formal model of this scenario requires receivers to distinguish three states: $\ket 0$ and $\ket 1$, which span the subspace wherein the sender can encode a one-qubit message, plus a void state, denoted $\ket v$, which represents not having received a message at all. 
The mapping from the sender's input to the final state of the receivers is given by
\begin{align}
\ket\psi_A  \rightarrow \ket{0}_A \otimes \frac{1}{\sqrt{N}} \left(\sum_{i=1}^N \ket{v}_1 ... \ket\psi_i ... \ket{v}_N\right) 
\end{align}
The partial state of a single receiver, which is the output of the channel $\Phi$, is obtained by tracing out all other receivers:
\begin{align}\label{eq:querasurechannel}
\Phi\left(\ketbra\psi\psi\right) =\frac{1}{N}\ketbra\psi\psi +\frac{N-1}{N}\ketbra{v}{v}.
\end{align}

The channel in \eqref{eq:querasurechannel} is known as the quantum erasure channel \cite{grassl_codes_1997,hayashi_quantum_2017a,caruso_quantum_2014,wilde2013quantum}, since it effectively erases the input with probability $p=(N-1)/N$, and it is known to have zero quantum capacity for $N\geq 2$,  assuming no classical side-channels \cite{bennett_capacities_1997,hayashi_quantum_2017a,caruso_quantum_2014,wilde2013quantum}.

Channels of this form are often encountered in quantum optics, where photons carrying quantum information may be lost on the way. Nonetheless, these systems are considered to allow for quantum communication, i.e., to offer a channel with non-zero quantum capacity. The key in such scenarios is  that a failure of the channel (such as the loss of the photon) is heralded, which allows the receiver to tally only those runs in which a message was actually received. Mathematically,  this corresponds to the
receivers post-selecting on the outcome associated with the projector $\id-\ket{v}\bra{v}$. Under this condition, 
the component describing the information that a single Bob receives \textendash{} by the photons that actually reach him \textendash{} is the identity channel, which has unit quantum capacity.

\subsubsection{Delocalizing quantum information among receivers}

We now exploit the converse of the first observation: 
if quantum information is sent using symmetrically emitted signals and there are no additional degrees of freedom that encode where the transmission is going, then the information cannot be recovered coherently by a single, localized receiver. Instead, receivers must cooperate in order to obtain a channel with non-zero quantum capacity.

This class of scenarios is relevant from a fundamental point of view since it addresses the question how quantum information can be delocalized in a quantum field
(cf. \cite{hayden_summoning_2012-1,adlam_quantum_2015}). It could also be of interest for future implementations of tasks such as quantum bit commitment or quantum secret sharing \cite{kent_unconditionally_1999,hillery_quantum_1999,gottesman_theory_2000}.

A minimal formal model of this scenario, which corresponds to the scenario discussed in Section \ref{sec:delocalizingcavity} (see Equation \eqref{eq:tworeceiverfinalstate}), begins with a generic initial state of the sender,  
\begin{align}
\ket\psi_A&=c_0\ket0_A +c_1 \ket1_A 
\end{align}
which is mapped to multipartite entangled states of all receivers,  
\begin{align} \label{eq:symmetric_GHZ}
\ket\psi_A\to c_0\ket0^{\otimes N}+c_1\ket1^{\otimes N}.
\end{align}
The partial state of any single receiver is then
\begin{align}
 \Phi\left(\ketbra\psi\psi\right) = \left|c_0\right|^2 \ketbra00 + \left|c_1\right|^2 \ketbra11,
\end{align}
which only contains information about a single expectation value, $\bra\psi \sigma_Z\ket\psi=\left|c_0\right|^2-\left|c_1\right|^2$, but not about coherences between the corresponding basis states. In order to obtain coherent information about the sender's initial quantum state, the receivers need to bring their shares of the output state together.
This entails a lower bound on the time it takes the two receivers to retrieve the sender's state, due to relativistic constraints arising from the separation between the receivers combined with the speed of light as an upper bound in information transfer.

\section{Multi-mode Coherent States and Displacement Operators}\label{app:coherent}
We collect a few formulae and properties of multi-mode coherent states following the conventions and definitions of \cite{barnett_methods_1997}.
Continuous multi-mode displacement operators are defined as
\begin{align}
D_{\alpha_1}=\exp\left(\integral{k}{-\infty}\infty \alpha_1(k) a^\dagger_k- \alpha_1(k)^* a_k\right),
\end{align}
and they displace the vacuum into the coherent state
\begin{align}
\ket{\alpha_1}=D_{\alpha_1}\ket0.
\end{align}

The scalar product of two coherent states is
\begin{align}\label{eq:coherentstateproduct}
\braket{\alpha_1}{\alpha_2}&=\ee{-\frac12\integral{k}{}{}\left( |\alpha_1(k)|^2 +|\alpha_2(k)|^2-2 \alpha_2(k) \alpha_1^*(k)\right)},
\end{align}
in particular
\begin{align}
\braket{\alpha_1}0=\ee{-\frac12\|\alpha_1\|^2}
\end{align}
and
\begin{align}\label{eq:coherentoverlapepsilon}
\braket{\alpha+\epsilon}{\alpha}&= \ee{-\frac12 \integral{k}{}{} |\epsilon(k)|^2} \ee{\ii\Im [\integral{k}{}{} \alpha(k) \epsilon(k)^*]}\nn 
&= \ee{-\|\epsilon\|^2/2} \ee{\ii \varphi(\alpha,\epsilon)}
\end{align}
where we introduced the notation
\begin{align}
\|\epsilon\|^2=\integral{k}{}{} \left|\epsilon(k)\right|^2 \label{eq:displacementnorm}\\ 
\varphi(\alpha,\epsilon)=\Im\integral{k}{}{} \alpha(k) \epsilon(k)^* .
\end{align}

One can see that the coherent states $\ket{\alpha}$ and $\ket{\alpha+\epsilon}$ have a large overlap  
when $\| \epsilon\|^2$ is small. Consequently, an equally weighted superposition of the two can vanish depending on the relative phase of the terms; in particular
\begin{align}
&\|\ket{\alpha}\pm\ee{\ii\varphi(\epsilon,\alpha)}\ket{\alpha+\epsilon}\|^2\nn 
&\qquad =2\left(1\pm\ee{-\|\epsilon\|^2/2}\cos\left(2\varphi(\epsilon,\alpha)\right) \right).
\end{align}

The composition of two displacement operators is
\begin{align}\label{eq:displacementcomposition}
D_{\alpha_1} D_{\alpha_2}=\ee{\ii\varphi(\alpha_1,\alpha_2)} D(\alpha_1+\alpha_2),
\end{align}
so that $ D_{\alpha_1} \ket{\alpha_2}= \ee{\ii\varphi(\alpha_1,\alpha_2)} \ket{\alpha_1+\alpha_2}$.
This follows from the Baker-Campbell-Hausdorff formula,
\begin{align}
\exp( A+B)=\exp( A)\exp(  B) \exp\left(-\frac12  \comm{A}{B}\right),
\end{align}
which holds when $\comm{A}{\comm{A}{B}}=\comm{B}{\comm{A}{B}}=0$.
One can see that the complex phase factor $\varphi(\alpha_1,\alpha_2)$ is actually related to the commutator of the field operators generating the displacement.
Writing
\begin{align}
D_{\alpha_1}=\exp(-\ii \Phi_A), \qquad D_{\alpha_2}=\exp(-\ii \Phi_B),
\end{align}
with $\Phi_A=\ii\left(\integral{k}{}{}\alpha_1 a_k^\dagger-\alpha_1^* a_k\right)$ and $\Phi_B$ analogously, we have
\begin{align}\label{eq:phifromcomm}
\comm{\Phi_A}{\Phi_B}&=2 \ii\, \Im \integral{k}{}{} \alpha_1^* \alpha_2 \nn 
&= 2\ii\varphi(\alpha_2,\alpha_1)=-2\ii\varphi(\alpha_1,\alpha_2).
\end{align}
In particular, the phase $\varphi(\alpha_1,\alpha_2)$ vanishes if the field observables commute, $\comm{\Phi_A}{\Phi_B}=0$, in which case the two displacement operators commute as well, since
\begin{align}\label{eq:commD}
\comm{D_{\alpha_1}}{D_{\alpha_2}}=2\ii\sin\left(\varphi(\alpha_1, \alpha_2) \right) D(\alpha_1+\alpha_2).
\end{align}

\section{Rewriting the coherent information between ancilla and field}\label{app:coherentinfo}

Let $A'$ denote an ancilla qubit, which is prepared in a maximally entangled state with Alice's detector,
\begin{align}\label{eq:bellstateinitial_appendix}
\ket{\psi}_{AA'}=\frac1{\sqrt2} \left(\ket{+X}_A\ket{+X}_{A'}-\ket{-X}_A\ket{-X}_{A'}\right).
\end{align}
After the two interactions, we obtain a state $\rho_{AA'F}=U \ket\psi_{AA'}\ket0_F \bra\psi_{AA'}\bra0_F U^\dagger$, which may generally contain entanglement between Alice, the ancilla, and the field degrees of freedom. 

We are interested in the coherent information between the ancilla $A'$ and the field $F$ in this state,
\begin{align}\label{eq:coherentinfoalicetofield}
I(A'>F)&=
S\left( \tr_{AA'} \rho_{AA'F} \right) 
- S\left( \tr_{A} \rho_{AA'F} \right), 
\end{align}
where $S(\rho)= -\tr \rho \log_2\rho$ denotes the von Neumann entropy of the state $\rho$ and, e.g., $\tr_A \rho$ denotes the partial trace over Alice's detector.
The evaluation of $I(A'>F)$ can be simplified by replacing the partial states above, which act on the infinite-dimensional Hilbert space of the field, with partial states of the two qubits only. 
This is possible because the  initial state $\ket{\psi_{AA'}}\ket0_F$, and consequently also the final state $\rho_{AA'F}$, are pure. Therefore,  the von Neumann entropies of complementary partial states are equal, e.g., $S\left(\tr_{AA'}\rho_{AA'F}\right)=S\left(\tr_{F}\rho_{AA'F}\right)$. This allows us to rewrite the relevant coherent information in terms of the partial state on Alice's detector and the ancilla, $\rho_{AA'}\equiv\tr_{F} \rho_{AA'F}$, as
\begin{align}
I(A'>F)&=
S\left(\rho_{AA'} \right) - S\left( \tr_{A'} \rho_{AA'} \right).
\end{align}

\section{Sensing the field displacement by coupling to the field amplitude}\label{app:amplitudecoupling}
Since Alice encoded her signal into $\pi_-$, the right-moving momentum of the field, and the amplitude and the momentum of the field are canonically conjugate to each other, it may appear natural to read out the signal by having Bob couple to the field amplitude. 
However, Bob cannot couple directly to the right-moving field amplitude $\phi_-(x,t)$ by itself, because it is not a localized field  observable \cite{jonsson_information_2016}. 
Instead, we should couple Bob to the full amplitude of the field, comprising left- and right-moving modes,
\begin{align}
\phi(x,t)=\integral{k}{-\infty}\infty \frac1{\sqrt{4\pi|k|}} \left( \ee{-\ii(|k| t-k x)} a_k + \ee{\ii(|k| t-k x)} a_k^\dagger\right),
\end{align}
in order to properly model his detector as being localized within the support of his profile function.

The interaction Hamiltonian generating $V^{(1)}$ by coupling to the field amplitude at time $t=t_2$ then reads
\begin{align}
\intHd{B}^{(1)}=\frac{\lambda_B}2 \delta(t-t_2)\left(\id-\sigma_X\right) \otimes\integral{x}{}{} g(x)\phi(x,t).
\end{align}
If we assume that Bob's profile is identical to Alice's but shifted to the right, 
$g(x)=f(x-t_2)$, then  $\tilde{g}(k)= \ee{-\ii k t_2} \tilde{f}(k)$ (compare \eqref{eq:profilefourier}). This yields
\begin{align}\label{eq:gamma}
\gamma_1(k)= \frac{-\ii\lambda_B \ee{2\ii\theta(-k) |k|t_2}}{\sqrt{4\pi|k|}} \tilde{f}(k)
\end{align}
for the resulting multi-mode displacement, where $\theta(k)$ denotes the Heaviside function. Consequently, 
\begin{align}
\varphi(\gamma_1,\alpha_1)=-\frac{\lambda_B\mu_A}{4\pi}\integral{k}0\infty |\tilde{f}(k)|^2.
\end{align}
Thus, the requirement of $\varphi(\gamma_1,\alpha_1)=\pi/2$  can be fulfilled by 
\begin{align}
\frac1{\lambda_B}=-\frac{\mu_A }{2\pi^2}\integral{k}0\infty |\tilde{f}(k)|^2.
\end{align}

It would seem that this choice of $\lambda_B$ would also allow us to satisfy the requirement that $\|\gamma_1\|^2=\integral{k}{}{} |\gamma_1(k)|^2$ be small, by choosing $\mu_A$ to be large. However, a problem arises due to the infrared (IR) divergence of massless fields in 1+1 dimensions.
In fact, in free Minkowski spacetime the norm  $\|\gamma_1\|^2$ of the displacement \eqref{eq:gamma} is IR-divergent. Therefore, an IR cutoff is required, which could be naturally introduced by considering the setup inside a cavity with vanishing Dirichlet boundary conditions. Here the field is expanded into discrete modes and, in particular, there is no zero-mode.  While this would render $\|\gamma_1\|^2$ finite, it might still lead to large values of $\|\gamma_1\|^2$ which decreases the fidelity of the state transfer.

\section{Lower Bound on Overlap with Target State}\label{app:overlap}

To obtain a lower bound on the overlap between Bob's final state and the ideal target state, we rewrite the final state of the system in the form 
\begin{align}
 VU\ket\zeta= \ket\psi_B\ket{v}_{A,F}+\ket{\psi^\perp}_B\ket{w}_{A,F}.
\end{align}
Here $\ket{\psi^\perp}= |x_-| \ee{\ii\arg x_+}\ket{+X}+|x_+|\ee{\ii\arg x_-}\ket{-X}$ is the detector state orthogonal to the target state $\ket\psi$.
We further split the states of Alice's detector and the field into
\begin{align}
    \ket{w}_{A,F}=\frac1{\sqrt2} \left(\ket{+Z}_A\ket{w^{(+)} }+\ket{-Z}_A \ket{w^{(-)}} \right)
\end{align}
and analogously for $\ket{v}_{A,F}$.

Fully expanded and exact, the final state of the detectors and the field for arbitrary intial states $\ket\psi_A$ of Alice reads
\begin{widetext}
\begin{align}
VU\ket \zeta&=\frac{-1}{2\sqrt2} \ket\psi \left[\ket{+Z}_A \left( |x_+|^2 \left(\ket{\alpha_2}+\ket{\gamma_1+\alpha_2}\right)+x_+^* x_-\left(\ket{\alpha_2-\alpha_1}-\ii\ket{\gamma_1+\alpha_2-\alpha_1}\right) \right.\right.\nn
&\qquad \left.\left.- x_-^* x_+ \left( \ii  \ket{\gamma_2+\alpha_2+\alpha_1} -\ket{\gamma_2+\gamma_1+\alpha_2+\alpha_1}\right) -| x_-|^2  \ii\left( \ket{\gamma_2+\alpha_2}+\ket{\gamma_2+\gamma_1+\alpha_2}\right) \right) \right.\nn
&\quad \left. +\ket{-Z}_A  \left( |x_+|^2 \left( \ket{0}+\ket{\gamma_1}\right) +x_+^* x_-\left(- \ket{-\alpha_1}+\ii\ket{\gamma_1-\alpha_1}\right)\right.\right.\nn
&\qquad\qquad\left.\left.- x_-^* x_+   \left( \ket{\gamma_2+\alpha_1} +\ii\ket{\gamma_2+\gamma_1+\alpha_1}\right) +|x_-|^2 \left( \ket{\gamma_2}+\ket{\gamma_2+\gamma_1}\right) \right) \right]\nn
&+\frac{-1}{2\sqrt2} \ket{\psi^\perp} \left[\ket{+Z}_A \left( |x_-|\,|x_+| \left(\ket{\alpha_2}+\ket{\gamma_1+\alpha_2}\right)+|x_-|\ee{-\ii\arg x_+}x_- \left(\ket{\alpha_2-\alpha_1}-\ii\ket{\gamma_1+\alpha_2-\alpha_1}\right) \right.\right.\nn
&\qquad \left.\left.+|x_+|\ee{-\ii\arg x_-} x_+ \left( \ii  \ket{\gamma_2+\alpha_2+\alpha_1} -\ket{\gamma_2+\gamma_1+\alpha_2+\alpha_1}\right) +|x_+| |x_-|  \ii\left( \ket{\gamma_2+\alpha_2}+\ket{\gamma_2+\gamma_1+\alpha_2}\right) \right) \right.\nn
&\quad \left. +\ket{-Z}_A  \left( |x_-|\,|x_+| \left( \ket{0}+\ket{\gamma_1}\right) +|x_-|\ee{-\ii\arg x_+} x_-\left(- \ket{-\alpha_1}+\ii\ket{\gamma_1-\alpha_1}\right)\right.\right.\nn
&\qquad\qquad\left.\left.+|x_+|\ee{-\ii\arg x_-} x_+   \left( \ket{\gamma_2+\alpha_1} +\ii\ket{\gamma_2+\gamma_1+\alpha_1}\right) -|x_+|\,|x_-| \left( \ket{\gamma_2}+\ket{\gamma_2+\gamma_1}\right) \right) \right].
\end{align}

From this we read off:
\begin{align}
\ket{w^{(+)}}
&= -\frac12\left( |x_-|\,|x_+| \left(\ket{s_1}+\ii\ket{s_2}\right) +|x_+|\ee{-\ii\arg x_-} x_+ \ket{r_2} +|x_-|\ee{-\ii\arg x_+}x_- \ket{r_1}  \right)\\
\ket{w^{(-)}}
&=-\frac12 \left( |x_-|\,|x_+| \left( \ket{s_3}-\ket{s_4}\right) +|x_-|\ee{-\ii\arg x_+} x_-\ket{r_3} +|x_+|\ee{-\ii\arg x_-} x_+   \ket{r_4}  \right)
\end{align}
\end{widetext}
where we defined
\begin{align}
&\ket{s_1}=\ket{\alpha_2}+\ket{\gamma_1+\alpha_2}\\
&\ket{s_2}=\ket{\gamma_2+\alpha_2}+\ket{\gamma_2+\gamma_1+\alpha_2}\\
&\ket{s_3}=\ket0+\ket{\gamma_1}\\
&\ket{s_4}=\ket{\gamma_2}+\ket{\gamma_2+\gamma_1}\\
&\ket{r_1}=\ket{\alpha_2-\alpha_1}-\ii\ket{\gamma_1+\alpha_2-\alpha_1}\\
&\ket{r_2}=\ii\ket{\gamma_2+\alpha_2+\alpha_1}-\ket{\gamma_2+\gamma_1+\alpha_2+\alpha_1}\\
&\ket{r_3}=-\ket{-\alpha_1}+\ii\ket{\gamma_1-\alpha_1}\\
&\ket{r_4}=\ket{\gamma_2+\alpha_1}+\ii\ket{\gamma_2+\gamma_1+\alpha_1}.
\end{align}
The norm of the $\ket{r_i}$ field states is upper bounded by the size of the displacement $\|\gamma_1\|^2$ as defined in \eqref{eq:displacementnorm}
\begin{align}
\braket{r_i}{r_i}=2(1-\ee{-\|\gamma_1\|^2/2}) 
<\|\gamma_1\|^2.
\end{align}
The $\ket{s_i}$ appear in pairwise superpositions. The combined norm of these pairs is bounded by
\begin{align}
&\|\ket{s_1}+\ii \ket{s_2}\|^2 =\|\ket{s_3} - \ket{s_4}\|^2 \nn 
&=4(1+\ee{-\|\gamma_1\|^2/2}) \nn* 
&\qquad -2  \left(2 \ee{-\|\gamma_2\|^2/2}+\ee{-\|\gamma_2+\gamma_1\|^2/2} +\ee{-\|\gamma_1-\gamma_2\|^2/2}\right)\nn*
&< 2 \underbrace{\left(\|\gamma_2\|^2-\|\gamma_1\|^2\right)}_{=0}+\underbrace{\|\gamma_2+\gamma_1\|^2+\|\gamma_1-\gamma_2\|^2 }_{=2\|\gamma_2\|^2+2 \|\gamma_1\|^2 = 4\|\gamma_1\|^2} \nn 
&\quad= 4\|\gamma_1\|^2
\end{align}
and we henceforth assume that $\|\gamma_2\|^2=\|\gamma_1\|^2$, since the two read-out couplings of Bob to which these displacements correspond are typically of the same strength. 

Then the norm of both the $\ket{w^{(+)}}$ and the $\ket{w^{(-)}}$ state is bounded by
\begin{align}
\braket{w^{(\pm)}}{w^{(\pm)}}&\leq \frac14  \left(|x_+|^2 |x_-|^2 \|\ket{s_1}+\ii\ket{s_2}\|^2 \right.\nn* 
&\left.\qquad +(|x_+|^4+|x_-|^4) \braket{r_i}{r_i}\right) \nn* 
<\frac12 \|\gamma_1\|^2
\end{align}
which, finally, yields
\begin{align}
\braket{w}{w}_{A,F}&=\frac12\left(\braket{w^{(+)}}{w^{(+)}}+\braket{w^{(-)}}{w^{(-)}}\right) \nn 
&<\frac{\|\gamma_1\|^2}2.
\end{align}
Thus, the overlap between Bob's exact final state $\rho_B=\tr_{A,F}\left( VU\ketbra\xi\xi U^\dagger V^\dagger \right)$, which in general is a mixed state, and the ideal pure target state $\ket\psi_B$ is lower bounded by
\begin{align}
 \tr \left[\rho_B \ketbra\psi\psi_B \right] = 1-\braket{w}{w}_{A,F}\geq 1-\frac12\|\gamma_1\|^2.
\end{align}

We note that this bound does not explicitly involve the strengths $\|\alpha_1\|^2$ and $\|\alpha_2\|^2$ of the initial displacements by Alice, which, as we showed in Section \ref{sec:CapacityAliceToField}, need to be strong in order to allow coherent information transfer from Alice into the field in the first place.
However, the requirement for $\|\gamma_1\|^2 $ to be small implies a lower bound on $\|\alpha_1\|^2$, because we also require that $\varphi(\gamma_1,\alpha_1)=\varphi(\gamma_2,\alpha_2)=\pi/2$: since
\begin{align}
0&\leq\integral{k}{}{} |\gamma_1(k)-\ii \alpha_1(k)|^2\nn 
&=\integral{k}{}{}|\gamma_1(k)|^2 + |\alpha_1(k)|^2 -2\underbrace{\Im \gamma_1(k) \alpha_1(k)^*}_{\varphi(\gamma_1,\alpha_1)}, 
\end{align}
the initial displacement need  to be at least as large as
\begin{align}
 \|\alpha_1\|^2 \geq \pi -\|\gamma_1\|^2, \qquad \|\alpha_2\|^2 \geq \pi -\|\gamma_2\|^2.
\end{align}

\bibliographystyle{unsrtnat}
\bibliography{krae_refs_Rob,krae_refs_Katja} 

\begin{thebibliography}{51}
\providecommand{\natexlab}[1]{#1}
\providecommand{\url}[1]{\texttt{#1}}
\expandafter\ifx\csname urlstyle\endcsname\relax
  \providecommand{\doi}[1]{doi: #1}\else
  \providecommand{\doi}{doi: \begingroup \urlstyle{rm}\Url}\fi

\bibitem[Rideout et~al.(2012)Rideout, Jennewein, Amelino-Camelia, Demarie,
  Higgins, {Achim Kempf}, Kent, Laflamme, Ma, Mann, Mart{\'\i}n-Mart{\'\i}nez,
  Menicucci, Moffat, Simon, Sorkin, Smolin, and
  Terno]{rideout_fundamental_2012}
David Rideout, Thomas Jennewein, Giovanni Amelino-Camelia, Tommaso~F. Demarie,
  Brendon~L. Higgins, {Achim Kempf}, Adrian Kent, Raymond Laflamme, Xian Ma,
  Robert~B. Mann, Eduardo Mart{\'\i}n-Mart{\'\i}nez, Nicolas~C. Menicucci, John
  Moffat, Christoph Simon, Rafael Sorkin, Lee Smolin, and Daniel~R. Terno.
\newblock Fundamental quantum optics experiments conceivable with
  satellites\textemdash{}reaching relativistic distances and velocities.
\newblock \emph{Classical and Quantum Gravity}, 29\penalty0 (22):\penalty0
  224011, 2012.
\newblock ISSN 0264-9381.
\newblock \doi{10.1088/0264-9381/29/22/224011}.
\newblock URL \url{http://stacks.iop.org/0264-9381/29/i=22/a=224011}.

\bibitem[Ren et~al.(2017)Ren, Xu, Yong, Zhang, Liao, Yin, Liu, Cai, Yang, Li,
  Yang, Han, Yao, Li, Wu, Wan, Liu, Liu, Kuang, He, Shang, Guo, Zheng, Tian,
  Zhu, Liu, Lu, Shu, Chen, Peng, Wang, and Pan]{ren_ground-to-satellite_2017}
Ji-Gang Ren, Ping Xu, Hai-Lin Yong, Liang Zhang, Sheng-Kai Liao, Juan Yin,
  Wei-Yue Liu, Wen-Qi Cai, Meng Yang, Li~Li, Kui-Xing Yang, Xuan Han,
  Yong-Qiang Yao, Ji~Li, Hai-Yan Wu, Song Wan, Lei Liu, Ding-Quan Liu, Yao-Wu
  Kuang, Zhi-Ping He, Peng Shang, Cheng Guo, Ru-Hua Zheng, Kai Tian, Zhen-Cai
  Zhu, Nai-Le Liu, Chao-Yang Lu, Rong Shu, Yu-Ao Chen, Cheng-Zhi Peng, Jian-Yu
  Wang, and Jian-Wei Pan.
\newblock Ground-to-satellite quantum teleportation.
\newblock \emph{Nature}, 549:\penalty0 70 EP --, 08 2017.
\newblock URL \url{http://dx.doi.org/10.1038/nature23675}.

\bibitem[Bradler(2011)]{bradler_infinite_2009}
K.~Bradler.
\newblock An infinite sequence of additive channels: The classical capacity of
  cloning channels.
\newblock \emph{IEEE Transactions on Information Theory}, 57\penalty0
  (8):\penalty0 5497--5503, Aug 2011.
\newblock ISSN 0018-9448.
\newblock \doi{10.1109/TIT.2011.2158896}.

\bibitem[Br{\'a}dler et~al.(2009)Br{\'a}dler, Hayden, and
  Panangaden]{bradler_private_2009}
Kamil Br{\'a}dler, Patrick Hayden, and Prakash Panangaden.
\newblock Private information via the {{Unruh}} effect.
\newblock \emph{Journal of High Energy Physics}, 2009\penalty0 (08):\penalty0
  074, 2009.
\newblock ISSN 1126-6708.
\newblock \doi{10.1088/1126-6708/2009/08/074}.
\newblock URL \url{http://stacks.iop.org/1126-6708/2009/i=08/a=074}.

\bibitem[Cliche and Kempf(2010)]{cliche_relativistic_2010}
M.~Cliche and A.~Kempf.
\newblock Relativistic quantum channel of communication through field quanta.
\newblock \emph{Physical Review A}, 81\penalty0 (1):\penalty0 012330, January
  2010.
\newblock \doi{10.1103/PhysRevA.81.012330}.
\newblock URL \url{http://link.aps.org/doi/10.1103/PhysRevA.81.012330}.

\bibitem[Bradler et~al.(2012)Bradler, Hayden, and
  Panangaden]{bradler_quantum_2012}
Kamil Bradler, Patrick Hayden, and Prakash Panangaden.
\newblock Quantum {{Communication}} in {{Rindler Spacetime}}.
\newblock \emph{Communications in Mathematical Physics}, 312\penalty0
  (2):\penalty0 361--398, June 2012.
\newblock ISSN 0010-3616, 1432-0916.
\newblock \doi{10.1007/s00220-012-1476-1}.
\newblock URL \url{http://arxiv.org/abs/1007.0997}.

\bibitem[Downes et~al.(2013)Downes, Ralph, and Walk]{downes_quantum_2013}
T.~G. Downes, T.~C. Ralph, and N.~Walk.
\newblock Quantum communication with an accelerated partner.
\newblock \emph{Physical Review A}, 87\penalty0 (1):\penalty0 012327, January
  2013.
\newblock \doi{10.1103/PhysRevA.87.012327}.
\newblock URL \url{http://link.aps.org/doi/10.1103/PhysRevA.87.012327}.

\bibitem[Bruschi et~al.(2014)Bruschi, Ralph, Fuentes, Jennewein, and
  Razavi]{bruschi_spacetime_2014}
David~Edward Bruschi, Timothy~C. Ralph, Ivette Fuentes, Thomas Jennewein, and
  Mohsen Razavi.
\newblock Spacetime effects on satellite-based quantum communications.
\newblock \emph{Physical Review D}, 90\penalty0 (4):\penalty0 045041, August
  2014.
\newblock \doi{10.1103/PhysRevD.90.045041}.
\newblock URL \url{https://link.aps.org/doi/10.1103/PhysRevD.90.045041}.

\bibitem[Landulfo(2016)]{landulfo_nonperturbative_2016}
Andr{\'e} G.~S. Landulfo.
\newblock Nonperturbative approach to relativistic quantum communication
  channels.
\newblock \emph{Physical Review D}, 93\penalty0 (10):\penalty0 104019, May
  2016.
\newblock \doi{10.1103/PhysRevD.93.104019}.
\newblock URL \url{http://link.aps.org/doi/10.1103/PhysRevD.93.104019}.

\bibitem[Jonsson(2017)]{jonsson_quantum_2017}
Robert~H. Jonsson.
\newblock Quantum signaling in relativistic motion and across acceleration
  horizons.
\newblock \emph{Journal of Physics A: Mathematical and Theoretical},
  50\penalty0 (35):\penalty0 355401, 2017.
\newblock ISSN 1751-8121.
\newblock \doi{10.1088/1751-8121/aa7d3c}.
\newblock URL \url{http://stacks.iop.org/1751-8121/50/i=35/a=355401}.

\bibitem[Gianfelici and Mancini(2017)]{gianfelici_quantum_2017}
Giulio Gianfelici and Stefano Mancini.
\newblock Quantum channels from reflections on moving mirrors.
\newblock \emph{Scientific Reports}, 7\penalty0 (1):\penalty0 15747, November
  2017.
\newblock ISSN 2045-2322.
\newblock \doi{10.1038/s41598-017-15578-0}.
\newblock URL \url{https://www.nature.com/articles/s41598-017-15578-0}.

\bibitem[Lloyd(1997)]{lloyd1997capacity}
Seth Lloyd.
\newblock Capacity of the noisy quantum channel.
\newblock \emph{Phys. Rev. A}, 55:\penalty0 1613--1622, Mar 1997.
\newblock \doi{10.1103/PhysRevA.55.1613}.
\newblock URL \url{https://link.aps.org/doi/10.1103/PhysRevA.55.1613}.

\bibitem[Kretschmann and Werner(2004)]{kretschmann2004capacity}
Dennis Kretschmann and Reinhard~F Werner.
\newblock Tema con variazioni : quantum channel capacity.
\newblock \emph{New Journal of Physics}, 6\penalty0 (1):\penalty0 26, 2004.
\newblock URL \url{http://stacks.iop.org/1367-2630/6/i=1/a=026}.

\bibitem[Bennett et~al.(1996)Bennett, DiVincenzo, Smolin, and
  Wootters]{bennett1996entangQEC}
Charles~H. Bennett, David~P. DiVincenzo, John~A. Smolin, and William~K.
  Wootters.
\newblock Mixed-state entanglement and quantum error correction.
\newblock \emph{Phys. Rev. A}, 54:\penalty0 3824--3851, Nov 1996.
\newblock \doi{10.1103/PhysRevA.54.3824}.
\newblock URL \url{https://link.aps.org/doi/10.1103/PhysRevA.54.3824}.

\bibitem[Kent(1999)]{kent_unconditionally_1999}
Adrian Kent.
\newblock Unconditionally {{Secure Bit Commitment}}.
\newblock \emph{Physical Review Letters}, 83\penalty0 (7):\penalty0 1447--1450,
  August 1999.
\newblock \doi{10.1103/PhysRevLett.83.1447}.
\newblock URL \url{http://link.aps.org/doi/10.1103/PhysRevLett.83.1447}.

\bibitem[Hillery et~al.(1999)Hillery, Bu{\v z}ek, and
  Berthiaume]{hillery_quantum_1999}
Mark Hillery, Vladim{\'\i}r Bu{\v z}ek, and Andr{\'e} Berthiaume.
\newblock Quantum secret sharing.
\newblock \emph{Physical Review A}, 59\penalty0 (3):\penalty0 1829--1834, March
  1999.
\newblock \doi{10.1103/PhysRevA.59.1829}.
\newblock URL \url{https://link.aps.org/doi/10.1103/PhysRevA.59.1829}.

\bibitem[Gottesman(2000)]{gottesman_theory_2000}
Daniel Gottesman.
\newblock Theory of quantum secret sharing.
\newblock \emph{Physical Review A}, 61\penalty0 (4):\penalty0 042311, March
  2000.
\newblock \doi{10.1103/PhysRevA.61.042311}.
\newblock URL \url{https://link.aps.org/doi/10.1103/PhysRevA.61.042311}.

\bibitem[Adlam and Kent(2015)]{adlam_device-independent_2015}
Emily Adlam and Adrian Kent.
\newblock Device-independent relativistic quantum bit commitment.
\newblock \emph{Physical Review A}, 92\penalty0 (2):\penalty0 022315, August
  2015.
\newblock \doi{10.1103/PhysRevA.92.022315}.
\newblock URL \url{https://link.aps.org/doi/10.1103/PhysRevA.92.022315}.

\bibitem[Cirac et~al.(1997)Cirac, Zoller, Kimble, and
  Mabuchi]{cirac_quantum_1997}
J.~I. Cirac, P.~Zoller, H.~J. Kimble, and H.~Mabuchi.
\newblock Quantum {{State Transfer}} and {{Entanglement Distribution}} among
  {{Distant Nodes}} in a {{Quantum Network}}.
\newblock \emph{Physical Review Letters}, 78\penalty0 (16):\penalty0
  3221--3224, April 1997.
\newblock \doi{10.1103/PhysRevLett.78.3221}.
\newblock URL \url{http://link.aps.org/doi/10.1103/PhysRevLett.78.3221}.

\bibitem[Mart{\'\i}n-Mart{\'\i}nez(2015)]{martin-martinez_causality_2015}
Eduardo Mart{\'\i}n-Mart{\'\i}nez.
\newblock Causality issues of particle detector models in {{QFT}} and quantum
  optics.
\newblock \emph{Physical Review D}, 92\penalty0 (10):\penalty0 104019, November
  2015.
\newblock \doi{10.1103/PhysRevD.92.104019}.
\newblock URL \url{http://link.aps.org/doi/10.1103/PhysRevD.92.104019}.

\bibitem[Jonsson et~al.(2015)Jonsson, Mart{\'\i}n-Mart{\'\i}nez, and
  Kempf]{jonsson_information_2015}
Robert~H. Jonsson, Eduardo Mart{\'\i}n-Mart{\'\i}nez, and Achim Kempf.
\newblock Information {{Transmission Without Energy Exchange}}.
\newblock \emph{Physical Review Letters}, 114\penalty0 (11):\penalty0 110505,
  March 2015.
\newblock \doi{10.1103/PhysRevLett.114.110505}.
\newblock URL \url{http://link.aps.org/doi/10.1103/PhysRevLett.114.110505}.

\bibitem[Jonsson(2016)]{jonsson_information_2016}
Robert~H. Jonsson.
\newblock Information travels in massless fields in 1+1 dimensions where energy
  cannot.
\newblock \emph{Journal of Physics A: Mathematical and Theoretical},
  49\penalty0 (44):\penalty0 445402, 2016.
\newblock ISSN 1751-8121.
\newblock \doi{10.1088/1751-8113/49/44/445402}.
\newblock URL \url{http://stacks.iop.org/1751-8121/49/i=44/a=445402}.

\bibitem[Leghtas et~al.(2013)Leghtas, Kirchmair, Vlastakis, Devoret,
  Schoelkopf, and Mirrahimi]{leghtas_deterministic_2013}
Zaki Leghtas, Gerhard Kirchmair, Brian Vlastakis, Michel~H. Devoret, Robert~J.
  Schoelkopf, and Mazyar Mirrahimi.
\newblock Deterministic protocol for mapping a qubit to coherent state
  superpositions in a cavity.
\newblock \emph{Physical Review A}, 87\penalty0 (4):\penalty0 042315, April
  2013.
\newblock \doi{10.1103/PhysRevA.87.042315}.
\newblock URL \url{http://link.aps.org/doi/10.1103/PhysRevA.87.042315}.

\bibitem[Blais et~al.(2004)Blais, Huang, Wallraff, Girvin, and
  Schoelkopf]{blais_cavity_2004}
Alexandre Blais, Ren-Shou Huang, Andreas Wallraff, S.~M. Girvin, and R.~J.
  Schoelkopf.
\newblock Cavity quantum electrodynamics for superconducting electrical
  circuits: {{An}} architecture for quantum computation.
\newblock \emph{Physical Review A}, 69\penalty0 (6):\penalty0 062320, June
  2004.
\newblock \doi{10.1103/PhysRevA.69.062320}.
\newblock URL \url{https://link.aps.org/doi/10.1103/PhysRevA.69.062320}.

\bibitem[McKay et~al.(2017)McKay, Lupascu, and
  Mart{\'\i}n-Mart{\'\i}nez]{mckay_finite_2017}
Emma McKay, Adrian Lupascu, and Eduardo Mart{\'\i}n-Mart{\'\i}nez.
\newblock Finite sizes and smooth cutoffs in superconducting circuits.
\newblock \emph{Physical Review A}, 96\penalty0 (5):\penalty0 052325, November
  2017.
\newblock \doi{10.1103/PhysRevA.96.052325}.
\newblock URL \url{https://link.aps.org/doi/10.1103/PhysRevA.96.052325}.

\bibitem[Forn-D{\'\i}az et~al.(2017)Forn-D{\'\i}az, Garc{\'\i}a-Ripoll,
  Peropadre, Orgiazzi, Yurtalan, Belyansky, Wilson, and
  Lupascu]{forn-diaz_ultrastrong_2017}
P.~Forn-D{\'\i}az, J.~J. Garc{\'\i}a-Ripoll, B.~Peropadre, J.-L. Orgiazzi,
  M.~A. Yurtalan, R.~Belyansky, C.~M. Wilson, and A.~Lupascu.
\newblock Ultrastrong coupling of a single artificial atom to an
  electromagnetic continuum in the nonperturbative regime.
\newblock \emph{Nature Physics}, 13\penalty0 (1):\penalty0 39, January 2017.
\newblock ISSN 1745-2481.
\newblock \doi{10.1038/nphys3905}.
\newblock URL \url{https://www.nature.com/articles/nphys3905}.

\bibitem[Johansson et~al.(2010)Johansson, Johansson, Wilson, and
  Nori]{johansson_dynamical_2010}
J.~R. Johansson, G.~Johansson, C.~M. Wilson, and Franco Nori.
\newblock Dynamical {{Casimir}} effect in superconducting microwave circuits.
\newblock \emph{Physical Review A}, 82\penalty0 (5):\penalty0 052509, November
  2010.
\newblock \doi{10.1103/PhysRevA.82.052509}.
\newblock URL \url{http://link.aps.org/doi/10.1103/PhysRevA.82.052509}.

\bibitem[Wilson et~al.(2011)Wilson, Johansson, Pourkabirian, Simoen, Johansson,
  Duty, Nori, and Delsing]{wilson_observation_2011}
C.~M. Wilson, G.~Johansson, A.~Pourkabirian, M.~Simoen, J.~R. Johansson,
  T.~Duty, F.~Nori, and P.~Delsing.
\newblock Observation of the dynamical {{Casimir}} effect in a superconducting
  circuit.
\newblock \emph{Nature}, 479\penalty0 (7373):\penalty0 376--379, November 2011.
\newblock ISSN 0028-0836.
\newblock \doi{10.1038/nature10561}.
\newblock URL
  \url{http://www.nature.com/nature/journal/v479/n7373/full/nature10561.html}.

\bibitem[DeWitt(1979)]{dewitt_quantum_1979}
B.~S. DeWitt.
\newblock Quantum gravity: The new synthesis.
\newblock In Stephen Hawking and W.~Israel, editors, \emph{General Relativity :
  An {{Einstein}} Centenary Survey}, page 680. {Cambridge University Press},
  Cambridge Eng; New York, 1979.
\newblock ISBN 978-0-521-22285-3.

\bibitem[Mart{\'\i}n-Mart{\'\i}nez et~al.(2013)Mart{\'\i}n-Mart{\'\i}nez,
  Montero, and {del Rey}]{martin-martinez_wavepacket_2013}
Eduardo Mart{\'\i}n-Mart{\'\i}nez, Miguel Montero, and Marco {del Rey}.
\newblock Wavepacket detection with the {{Unruh}}-{{DeWitt}} model.
\newblock \emph{Physical Review D}, 87\penalty0 (6):\penalty0 064038, March
  2013.
\newblock \doi{10.1103/PhysRevD.87.064038}.
\newblock URL \url{http://link.aps.org/doi/10.1103/PhysRevD.87.064038}.

\bibitem[Alhambra et~al.(2014)Alhambra, Kempf, and
  Mart{\'\i}n-Mart{\'\i}nez]{alhambra_casimir_2014}
{\'A}lvaro~M. Alhambra, Achim Kempf, and Eduardo Mart{\'\i}n-Mart{\'\i}nez.
\newblock Casimir forces on atoms in optical cavities.
\newblock \emph{Physical Review A}, 89\penalty0 (3):\penalty0 033835, March
  2014.
\newblock \doi{10.1103/PhysRevA.89.033835}.
\newblock URL \url{http://link.aps.org/doi/10.1103/PhysRevA.89.033835}.

\bibitem[Pozas-Kerstjens and
  Mart{\'\i}n-Mart{\'\i}nez(2016)]{pozas-kerstjens_entanglement_2016}
Alejandro Pozas-Kerstjens and Eduardo Mart{\'\i}n-Mart{\'\i}nez.
\newblock Entanglement harvesting from the electromagnetic vacuum with
  hydrogenlike atoms.
\newblock \emph{Physical Review D}, 94\penalty0 (6):\penalty0 064074, September
  2016.
\newblock \doi{10.1103/PhysRevD.94.064074}.
\newblock URL \url{https://link.aps.org/doi/10.1103/PhysRevD.94.064074}.

\bibitem[Hotta(2008)]{hotta_quantum_2008}
Masahiro Hotta.
\newblock Quantum measurement information as a key to energy extraction from
  local vacuums.
\newblock \emph{Physical Review D}, 78\penalty0 (4):\penalty0 045006, August
  2008.
\newblock \doi{10.1103/PhysRevD.78.045006}.
\newblock URL \url{http://link.aps.org/doi/10.1103/PhysRevD.78.045006}.

\bibitem[Simidzija and
  Mart\'{\i}n-Mart\'{\i}nez(2017)]{simidzija_non-perturbative_2017}
Petar Simidzija and Eduardo Mart\'{\i}n-Mart\'{\i}nez.
\newblock Nonperturbative analysis of entanglement harvesting from coherent
  field states.
\newblock \emph{Phys. Rev. D}, 96:\penalty0 065008, Sep 2017.
\newblock \doi{10.1103/PhysRevD.96.065008}.
\newblock URL \url{https://link.aps.org/doi/10.1103/PhysRevD.96.065008}.

\bibitem[Shor(2002)]{shor2002capacity}
Peter~W. Shor.
\newblock The quantum channel capacity and coherent information.
\newblock In \emph{Lecture Notes, MSRI Workshop on Quantum Computation}. 2002.

\bibitem[Devetak(2005)]{devetak2005private}
Igor Devetak.
\newblock The private classical capacity and quantum capacity of a quantum
  channel.
\newblock \emph{IEEE Transactions on Information Theory}, 51\penalty0
  (1):\penalty0 44--55, 2005.

\bibitem[Munro et~al.(2000)Munro, Milburn, and Sanders]{munro_entangled_2000}
W.~J. Munro, G.~J. Milburn, and B.~C. Sanders.
\newblock Entangled coherent-state qubits in an ion trap.
\newblock \emph{Physical Review A}, 62\penalty0 (5):\penalty0 052108, October
  2000.
\newblock \doi{10.1103/PhysRevA.62.052108}.
\newblock URL \url{http://link.aps.org/doi/10.1103/PhysRevA.62.052108}.

\bibitem[Horodecki et~al.(2006)Horodecki, Oppenheim, and
  Winter]{horodecki_quantum_2006}
Michal Horodecki, Jonathan Oppenheim, and Andreas Winter.
\newblock Quantum state merging and negative information.
\newblock \emph{Communications in Mathematical Physics}, 269\penalty0
  (1):\penalty0 107--136, November 2006.
\newblock ISSN 0010-3616, 1432-0916.
\newblock \doi{10.1007/s00220-006-0118-x}.
\newblock URL \url{http://arxiv.org/abs/quant-ph/0512247}.

\bibitem[Hayashi(2017)]{hayashi_quantum_2017a}
Masahito Hayashi.
\newblock \emph{Quantum {{Information Theory}}: {{Mathematical Foundation}}}.
\newblock Graduate Texts in Physics. {Springer-Verlag}, Berlin Heidelberg, 2
  edition, 2017.
\newblock ISBN 978-3-662-49723-4.
\newblock URL \url{//www.springer.com/gp/book/9783662497234}.

\bibitem[Caruso et~al.(2014)Caruso, Giovannetti, Lupo, and
  Mancini]{caruso_quantum_2014}
Filippo Caruso, Vittorio Giovannetti, Cosmo Lupo, and Stefano Mancini.
\newblock Quantum channels and memory effects.
\newblock \emph{Reviews of Modern Physics}, 86\penalty0 (4):\penalty0
  1203--1259, December 2014.
\newblock \doi{10.1103/RevModPhys.86.1203}.
\newblock URL \url{https://link.aps.org/doi/10.1103/RevModPhys.86.1203}.

\bibitem[Wilde(2013)]{wilde2013quantum}
Mark~M Wilde.
\newblock \emph{Quantum information theory}.
\newblock Cambridge University Press, 2013.

\bibitem[Pirandola et~al.(2017)Pirandola, Laurenza, Ottaviani, and
  Banchi]{pirandola_fundamental_2017}
Stefano Pirandola, Riccardo Laurenza, Carlo Ottaviani, and Leonardo Banchi.
\newblock Fundamental limits of repeaterless quantum communications.
\newblock \emph{Nature Communications}, 8:\penalty0 15043, April 2017.
\newblock ISSN 2041-1723.
\newblock \doi{10.1038/ncomms15043}.
\newblock URL \url{https://www.nature.com/articles/ncomms15043}.

\bibitem[Laurenza and Pirandola(2017)]{laurenza_general_2017}
Riccardo Laurenza and Stefano Pirandola.
\newblock General bounds for sender-receiver capacities in multipoint quantum
  communications.
\newblock \emph{Physical Review A}, 96\penalty0 (3):\penalty0 032318, September
  2017.
\newblock \doi{10.1103/PhysRevA.96.032318}.
\newblock URL \url{https://link.aps.org/doi/10.1103/PhysRevA.96.032318}.

\bibitem[Devetak and Shor(2005)]{devetak2005capacity}
Igor Devetak and Peter~W Shor.
\newblock The capacity of a quantum channel for simultaneous transmission of
  classical and quantum information.
\newblock \emph{Communications in Mathematical Physics}, 256\penalty0
  (2):\penalty0 287--303, 2005.

\bibitem[Holevo(2008)]{holevo2008entanglement}
Alexander~S Holevo.
\newblock Entanglement-breaking channels in infinite dimensions.
\newblock \emph{Problems of Information Transmission}, 44\penalty0
  (3):\penalty0 171--184, 2008.

\bibitem[Caruso and Giovannetti(2006)]{caruso2006degradability}
Filippo Caruso and Vittorio Giovannetti.
\newblock Degradability of bosonic gaussian channels.
\newblock \emph{Physical Review A}, 74\penalty0 (6):\penalty0 062307, 2006.

\bibitem[Grassl et~al.(1997)Grassl, Beth, and Pellizzari]{grassl_codes_1997}
M.~Grassl, Th. Beth, and T.~Pellizzari.
\newblock Codes for the quantum erasure channel.
\newblock \emph{Physical Review A}, 56\penalty0 (1):\penalty0 33--38, July
  1997.
\newblock \doi{10.1103/PhysRevA.56.33}.
\newblock URL \url{https://link.aps.org/doi/10.1103/PhysRevA.56.33}.

\bibitem[Bennett et~al.(1997)Bennett, DiVincenzo, and
  Smolin]{bennett_capacities_1997}
Charles~H. Bennett, David~P. DiVincenzo, and John~A. Smolin.
\newblock Capacities of {{Quantum Erasure Channels}}.
\newblock \emph{Physical Review Letters}, 78\penalty0 (16):\penalty0
  3217--3220, April 1997.
\newblock \doi{10.1103/PhysRevLett.78.3217}.
\newblock URL \url{https://link.aps.org/doi/10.1103/PhysRevLett.78.3217}.

\bibitem[Hayden and May(2012)]{hayden_summoning_2012-1}
Patrick Hayden and Alex May.
\newblock Summoning {{Information}} in {{Spacetime}}, or {{Where}} and {{When
  Can}} a {{Qubit Be}}?
\newblock \emph{arXiv:1210.0913 [gr-qc, physics:hep-th, physics:quant-ph]},
  October 2012.
\newblock URL \url{http://arxiv.org/abs/1210.0913}.

\bibitem[Adlam and Kent(2016)]{adlam_quantum_2015}
Emily Adlam and Adrian Kent.
\newblock Quantum paradox of choice: More freedom makes summoning a quantum
  state harder.
\newblock \emph{Phys. Rev. A}, 93:\penalty0 062327, Jun 2016.
\newblock \doi{10.1103/PhysRevA.93.062327}.
\newblock URL \url{https://link.aps.org/doi/10.1103/PhysRevA.93.062327}.

\bibitem[Barnett and Radmore(1997)]{barnett_methods_1997}
Stephen~M. Barnett and P.~M. Radmore.
\newblock \emph{Methods in {{Theoretical Quantum Optics}}}.
\newblock {Clarendon Press}, 1997.
\newblock ISBN 978-0-19-856362-4.

\end{thebibliography}

\end{document}